# A force-based beam element model based on the modified higher-order shear deformation theory for accurate analysis of FG beams

Wenxiong Li*, Huiyi Chen, Suiyin Chen*, Zhiwei Liu

*College of Water Conservancy and Civil Engineering, South China Agricultural University, Guangzhou 510642, China*

*Corresponding author. Email: leewenxiong@scau.edu.cn(W. Li), rinchan41@scau.edu.cn(S. Chen)

**ABSTRACT** This paper develops a modified higher-order shear deformation beam theory and a beam finite element model for the accurate analysis of functionally graded beams. The innovation of the modified higher-order shear deformation beam theory resides in the modified shear stiffness, which accurately delineates the relationship between shear deformation and shear force. Specifically, the differential equilibrium equation that describes the relationship between normal stress and shear stress is employed in this theory to derive the rational expression of transverse shear stress, thereby obtaining the accurate shear stiffness. Based on the modified theory, a force-based higher-order beam element model is firstly proposed, where the internal forces are regarded as the unknown fields. In the implementation of the force-based element, the internal forces are predefined from the closed-form solutions of the differential equilibrium equations of higher-order shear beam, and the generalized displacements can be expressed by the internal forces, according to the geometric and constitutive equations. Subsequently, the equation system of the beam element can be constructed based on the equilibrium conditions at the element boundaries and the compatibility condition within the element. Numerical examples are provided to illustrate the accuracy and effectiveness of the proposed beam element model.

## 1 Introduction

Functionally Graded (FG) materials, a unique class of composite materials, are distinguished by their property gradients along one or more dimensions. Owing to their exceptional mechanical properties, FG structures have found widespread use in a range of contemporary engineering applications, encompassing aerospace, marine, biomedical, and civil construction sectors [1]. FG materials are noted for their enhanced bond strength at layer interfaces, superior resistance to thermal stress, and an impressive strength-to-weight ratio. Given these attributes, the development of efficient and precise analysis models becomes crucial for accurately forecasting the behavior of FG structures under diverse loading conditions. This is a vital step towards harnessing the full potential of FG materials in various engineering applications.

For the analyses of FG beams, a range of beam theories, including Classical Beam Theory (CBT) [2-5], First-order Shear deformation Beam Theory (FSBT) [6-13], and Higher-order Shear deformation Beam Theory (HSBT) [14-26], have been employed. Beam models based on CBT, which neglect the effects of transverse shear deformation, are primarily suitable for slender beams. However, these models tend to overestimate stiffness and underestimate deflection for beams with a low slenderness ratio. Beam models based on FSBT do consider transverse shear deformation to a certain extent. Nevertheless, they operate under the assumption that the cross-section remains plane, necessitating a correction factor to adjust the shear stiffness. Yet, the determination of this shear correction factor for functionally graded beams remains a significant challenge. This highlights the need for further research and development in this area. Nguyen et al. [9] proposed an enhanced transverse stress stiffness for FSBT, taking into account the impact of transverse shear stress distribution based on the differential equilibrium equation. This advancement significantly improves the precision of beam models based on FSBT. Beam models based on HSBT employ a higher-order displacement function to delineate the distribution of longitudinal displacement through the thickness, thereby enhancing the accuracy in predicting strain and stress distributions. Typically, the higher-order displacement function in these beam models enables the derived transverse shear stress to more closely resemble the true distribution, such as zero transverse shear stress on the upper and lower boundaries. This allows for more accurate solutions without the need for a shear correction factor when the material properties exhibit smooth variation through the thickness. A multitude of studies have corroborated that HSBT-based beam models can yield more precise solutions [17, 19, 27-40]. Filippi et al. [14] evaluated various higher-order beam elements by means of the Carrera Unified Formulation (CUF). Vo et al. [41] devised a finite element based on Reddy-Bickford beam theory for the vibration and buckling analyses of FG sandwich beams, and scrutinized the effects of the power-law index, span-to-height ratio, core thickness, and boundary conditions. Incorporating a hyperbolic distribution of transverse shear stress, Nguyen et al. [42] developed a higher-order shear deformation beam model for the analysis of FG sandwich beams. They explored the effects of boundary conditions, power-law index, span-to-height ratio, and skin-core-skin thickness ratios on the critical buckling loads and natural frequencies. Belabed et al. proposed the new higher-order shear deformation theory for analyses of FG porous beams [21], FG sandwich beams [22] and bi-directional FG beams [24]. Mesbah et al. [23] presented the finite element model for free vibration and buckling behaviours of FG porous beams. Based on the higher-order shear deformation beam theory, the static bending response of rotating FG graphene platelet reinforced composite beams with geometrical imperfections in thermal mediums was investigated [26], and the simplified homogenization technique for nonlinear finite element analysis of in-plane loaded masonry walls was developed [25]. In addition to the analysis of beams, scholars have also developed corresponding first-order and higher-order shear deformation theories and conducted relevant research on functionally graded plates and shells. Based on the first-order shear deformation theory, recent achievements include static and buckling analyses of bi-directional FG plates [43], as well as free vibration analysis of FG sandwich plates with porosity [44]. Meanwhile, progress has been made in the nonlinear bending analysis of porous FG nanoplates [45]. Based on the higher-order shear deformation theory, geometric imperfection sensitivity on the vibration response of geometrically discontinuous bi-directional FG plates were investigated [46]. In recent years, some new finite element methods have been proposed based on higher-order shear deformation theory, such as the work for free vibration analysis of thick laminated composite shells [47] and the work for free vibration and bending analysis of porous FG sandwich shells [48]. In addition, significant achievements have been made in the analyses of FG nanoplates by combining nonlocal theory with the higher-order shear deformation theory [49, 50].

Despite certain advancements on the higher-order shear deformation beam theory and corresponding beam finite elements, the following two issues still need further improvement: the decrease in solution accuracy caused by inaccurate expression of transverse shear stress, and the solution errors caused by finite element discretization. The first issue arises from the traditional derivation method of beam elements, where the transverse shear stress is derived through geometric relations and constitutive equations, lacking assurance of the equilibrium differential equations. Due to not strictly satisfying the equilibrium relations, the difference between the transverse shear stress distribution derived from the traditional models and the true distribution is inevitable, and this difference is more significant in FG beams, especially in FG sandwich beams. In the context of FG sandwich beams, where material properties exhibit significant variations across the thickness, the transverse shear stress derived from the conventional higher-order shear deformation beam models exhibits a non-smooth distribution with abrupt changes at the interlayer junction, which will significantly differ from the true distribution of continuous smoothness, as shown in **Fig. 1**. The inaccurate distribution of transverse shear stress may adversely impact the accuracy of the solutions [51]. The second issue is actually the drawback of conventional finite element methods. Due to the use of simplified approximation functions to construct beam finite elements, discretization errors are inevitable. Therefore, mesh refinement is usually required to achieve the converged solutions. It should be noted that, for FG beams, the appropriate mesh refinement required to ensure convergence varies depending on the distribution of materials. In other words, it is difficult to accurately and quickly estimate an appropriate mesh refinement for convergent solutions, which affects the practical application of these beam elements. Meanwhile, the existence of discretization errors will affect the accuracy and credibility of the research conclusions on the behaviors of FG beams.

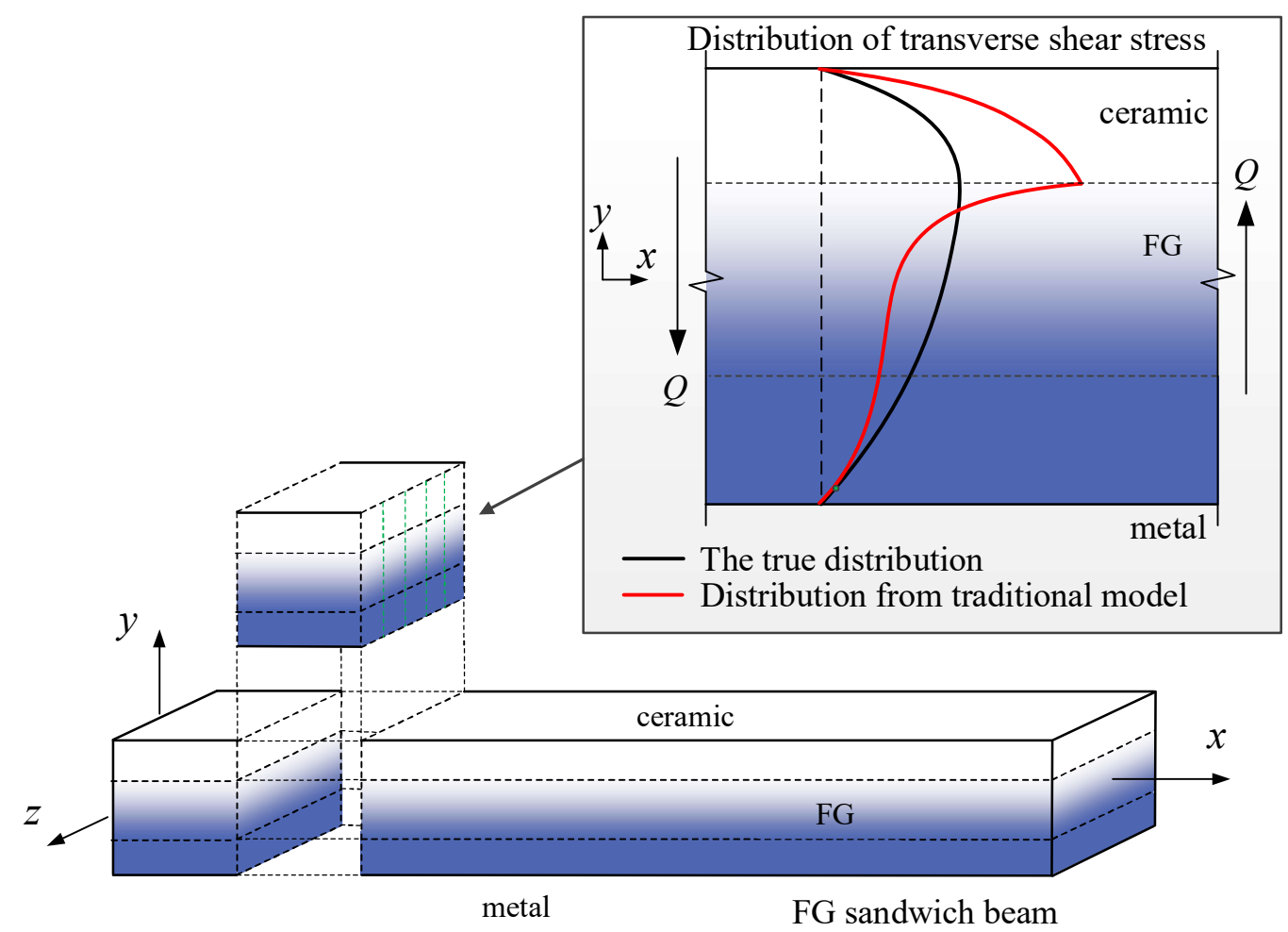

**Fig. 1**. Schematic diagram of transverse shear stress distribution.

To overcome the first problem, some modified theories and new beam element models are developed. Lezgy-Nazargah [52] developed a global-local shear deformation theory to improve the solution accuracy in shear stress. In this theory, a global kinematic describing the whole behaviour of the beam is put on local layer kinematic selected based on the layer-wise concepts. The continuity conditions of stress and displacement on the layer interfaces are used to reduce the number of unknown variables. Based on the global-local shear deformation theory, Lezgy-Nazargah [53] studied the coupling thermo-mechanical responses of 2D FG beams. Furthermore, the bending, vibration, and buckling behaviours of FG curved sandwich beams were investigated [54] and, subsequently, the stability and free vibration behavior of bidirectional FG sandwich beams were explored [55]. Li et al. [56] introduced a mixed higher-order shear beam element model to produce accurate transverse shear stress distributions. The central concept of this model is the incorporation of the differential equilibrium equation by establishing independent internal force fields, thereby enabling accurate prediction of the transverse shear stress distribution along the thickness direction. This mixed higher-order shear beam element model has also been utilized in the vibration analysis of FG sandwich beams [57]. An alternative approach to mitigate the aforementioned issue involves the use of a more rational higher-order displacement function. Ma [58] proposed a rational approach for determining the correct higher-order displacement function, which employs the equilibrium condition and considers different types of cross-sctions. In addition, Li et al. [59] proposed a material-based higher-order shear beam model, where the higher-order displacement function is constructed in accordance with the material distribution through the thickness and the differential equilibrium equation. In particular, the higher-order displacement function is characterized by a piecewise linear interpolation field and determined by ensuring the consistency of transverse shear stress distributions between Euler-Bernoulli beam theory and the higher-order shear beam theory. These new models can indeed derive accurate transverse shear stress distribution, thereby improving the accuracy of the solution. However, on the basis of the accurate expression of transverse shear stress, the theoretical research on higher-order beams is not sufficient. Firstly, there is no theoretical achievement on the distributions of internal forces in FG beams. In addition, these new beam finite element models are constructed based on the principle of strain energy variation or the principle of mixed variation, which cannot guarantee that the equilibrium equations are satisfied at both the element nodes and within the element. Therefore, the discretization error is still a problem that needs to be solved.

For the second problem, new methods of finite element construction should be developed to establish higher-order shear deformation beam elements that avoid discretization errors. In recent years, advancements have been made in the development of the exact finite element method, with notable progress reported in the areas of structural buckling analysis [60, 61] and structural vibration analysis [2, 62]. The methodology of the exact finite element involves the construction of high-precision finite element models utilizing interpolation functions derived from the closed-form solutions of the corresponding differential equilibrium equations. This approach has provided a framework for the development of high-precision higher-order shear beam element models. In static analysis, Ruocco and Reddy [63] discussed the closed-form solutions of the Reddy beam theory (a form of HSBT) and analyzed the bending behavior of straight and curved FG beams based on the derived closed-form solutions. Furthermore, they developed an exact beam finite element based on the closed-form solutions of generalized displacements, significantly advancing the development of higher-order beam finite element models with high-precision. Generally, for the same beam model, the differential equilibrium equations expressed in terms of internal forces has a lower order than those expressed in terms of generalized displacements. Therefore, compared with the generalized displacements, the closed-form solutions of the internal forces can be more easily derived from the differential equilibrium equations. In other words, it is more convenient to construct high-precision beam elements based on analytical internal force fields. In fact, some achievements have been made in the research on the force-based beam elements, such as Neuenhofer and Filippou [64, 65], Alemdar and White [66], Santos [67] and Li et al. [68], and their results indicate that the force-based beam elements typically have higher accuracy. The force-based finite element method has become a promising approach to solving the problem of discretization errors. However, to the authors' knowledge, the existing achievements on force-based beam elements are based on CBT or FSBT, while no relevant research on the development of the force-based beam element based on HSBT. Therefore, development of force-based higher-order shear deformable beam element models based on the analytical internal force fields is a worthwhile research topic. The authors believe that the force-based beam elements based on HSBT with accurate expression of transverse shear stress will exhibit extremely high-performance in behaviour analyses of FG beams.

To address the two aforementioned issues simultaneously, this paper develops a modified higher-order shear deformation beam theory and a beam finite element model for the accurate analysis of FG beams. The innovation of the proposed modified higher-order shear deformation beam theory lies in the modified shear stiffness that accurately describes the relationship between shear deformation and shear force. Unlike the conventional approach where the transverse shear stress is derived from geometric relations and constitutive equations, the differential equilibrium equation that describes the relationship between axial normal stress and transverse shear stress is employed to determine the accurate expression of transverse shear stress, thereby obtaining the accurate shear stiffness. Based on the modified beam theory, a force-based higher-order beam element model is firstly proposed. In contrast to the displacement-based beam elements and the mixed beam elements, such as [56, 59], the internal forces are treated as the unknown fields in the proposed beam element model. In the implementation of the force-based element, the internal forces are predefined from the closed-form solutions of the differential equilibrium equations of higher-order shear beam, and the generalized displacements can be expressed by the internal forces, according to the geometric relations and constitutive equations. Subsequently, the equation system of the beam element can be constructed based on the equilibrium conditions at the element boundaries and the compatibility condition within the element. Finally, numerical examples will be conducted to illustrate the accuracy and effectiveness of the proposed beam element model.

## 2 Formulation of the modified HSBT

### 2.1 Basic assumptions

As shown in **Fig. 2**, a beam of uniform rectangular cross-section, with width $b$, thickness $h$ and length $L$, is considered. The rectangular Cartesian coordinate axes with the $x$-axis along the geometric centroidal axis and the $y$-axis in the thickness direction is used to describe the positions, displacements and deformations of the beam. In the present study, the displacements and deformations in $x$-$y$ plane is considered and following assumptions are made:

(A1) The present study is a plane problem and allows for small strains and isotropic elasticity.

(A2) The beam is made of functional graded material and the material properties continuously vary in the thickness direction.

(A3) The beam undergos axial stretching deformation, bending deformation and transverse shear deformation. The beam's transverse stretching deformation is neglected, and hence the transverse normal strain and transverse normal stress are ignored.

(A4) The relationship between axial normal stress and transverse shear stress is established based on the equilibrium differential equation.

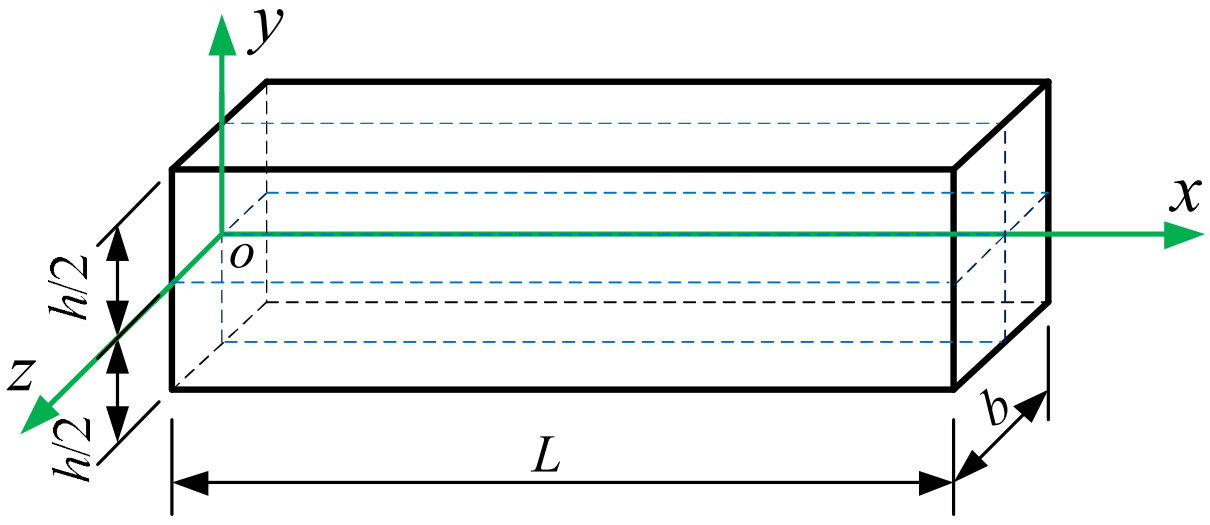

**Fig. 2**. Geometry and coordinate definition of a beam.

### 2.2 Displacements and strains

For a planar higher-order shear deformation beam model, the displacement fields can be expressed as [56]

$$\begin{aligned} u_x(x,y) &= u(x) - y\frac{\mathrm{d}w(x)}{\mathrm{d}x} + f(y)\left[\frac{\mathrm{d}w(x)}{\mathrm{d}x} - \theta(x)\right] \\ u_y(x,y) &= w(x) \end{aligned} \tag{1}$$

where $u(x)$ and $w(x)$ represent the axial and transverse displacements of any point on the beam's centre line respectively, $\theta(x)$ represents the rotation of the cross-section, $x \in [0, L]$ is the coordinate along the beam's length, and $y \in [-h/2, h/2]$ is the coordinate along the beam's thickness. Specifically, the positive directions of $u(x)$ and $w(x)$ are consistent with the *x*-axis and *y*-axis, respectively, and the positive direction of rotation is defined according to the right-hand rule with *z*-axis as the North pole. In Eq. (1), $f(y)$ represents a higher-order displacement function that varies along the direction of thickness. Various theories have been developed by choosing different forms of $f(y)$, according to the deviation in the shape of cross-section and the distribution of materials. For details on the forms of $f(y)$, readers are referred to Refs. [14, 16, 56, 58]. In the present study where the cross-section maintains rectangluar, the cubic form of $f(y)$ based on the classical Reddy beam theory [31, 63] is adopted. The expression of $f(y)$ is

$$f(y) = y\left(1 - \frac{4y^2}{3h^2}\right) \tag{2}$$

Based on the definition of displacement fields in Eq. (1), the expressions of non-zero strain components can be obtained as

$$\varepsilon_x(x, y) = \frac{\partial u_x(x, y)}{\partial x} = \varepsilon_0(x) + \left[y - f(y)\right]\kappa_w(x) + f(y)\kappa_\theta(x) \tag{3}$$

$$\gamma_{xy}(x, y) = \frac{\partial u_x(x, y)}{\partial y} + \frac{\partial u_y(x, y)}{\partial x} = f_{,y}(y)\gamma_0(x) \tag{4}$$

where $\varepsilon_0(x)$ denotes the axial strain of the centre line of the beam, $\kappa_w(x)$ represents the curvature associated with transverse displacement, $\kappa_\theta(x)$ represents the curvature related to the rotation of the cross-section, $\gamma_0(x)$ denotes the transverse shear deformation, and $f_{,y}(y) = \dfrac{\mathrm{d}f(y)}{\mathrm{d}y}$. The generalized strains $\varepsilon_0(x)$, $\kappa_w(x)$, $\kappa_\theta(x)$ and $\gamma_0(x)$ can be expressed by the displacements of the beam's axis as

$$\varepsilon_0(x) = u_{,x}(x) \tag{5}$$

$$\kappa_w(x) = -w_{,xx}(x) \tag{6}$$

$$\kappa_\theta(x) = -\theta_{,x}(x) \tag{7}$$

$$\gamma_0(x) = w_{,x}(x) - \theta(x) \tag{8}$$

where $(\cdot)_{,x} = \dfrac{\mathrm{d}(\cdot)}{\mathrm{d}x}$ and $(\cdot)_{,xx} = \dfrac{\mathrm{d}^2(\cdot)}{\mathrm{d}x^2}$ denotes the first-order and second-order derivatives with respect to $x$, respectively. Note that the positive directions of these generalized strains can also be determined from Eqs. (5)-(8).

For the sake of simplicity, the generalized strain components associated with the normal strain $\varepsilon_x(x, y)$ are represented by a vector as

$$\boldsymbol{\varepsilon}(x) = \left\{\varepsilon_0(x) \quad \kappa_w(x) \quad \kappa_\theta(x)\right\}^{\mathrm{T}} \tag{9}$$

Then, the normal strain in Eq. (3) can be expressed as

$$\varepsilon_x(x, y) = \mathbf{t}(y)\boldsymbol{\varepsilon}(x) \tag{10}$$

where

$$\mathbf{t}(y)=\begin{bmatrix}1 & y-f(y) & f(y)\end{bmatrix} \tag{11}$$

2.3 Constitutive relations

For the FG beams in which the material properties vary through the thickness, the relation between strains and stresses can be expressed as [16]

$$\sigma_x(x,y)=E(y)\varepsilon_x(x,y) \tag{12}$$

$$\tau_{xy}(x,y)=G(y)\gamma_{xy}(x,y) \tag{13}$$

where $\sigma_x(x,y)$ represents normal stress and $\tau_{xy}(x,y)$ represents transverse shear stress. Generally, the Young's modulus and shear modulus of the material maintain the following relationship

$$G(y)=\frac{E(y)}{2(1+v)} \tag{14}$$

where $v$ is Poisson's ratio of the material.

Based on the strain expression shown in Eqs. (3) and (4), the virtual strain energy of the beam can be rewritten as

$$\begin{aligned}\delta U&=\int_V\left[\sigma_x(x,y)\delta\varepsilon_x(x,y)+\tau_{xy}(x,y)\delta\gamma_{xy}(x,y)\right]\mathrm{d}V\\&=\int_0^L\left[N(x)\delta\varepsilon_0(x)+M_w(x)\delta\kappa_w(x)+M_\theta(x)\delta\kappa_\theta(x)+\hat{Q}_\theta(x)\delta\gamma_0(x)\right]\mathrm{d}x\end{aligned} \tag{15}$$

where $L$ represents the beam's length, $V$ represents the volume of the beam, and internal forces $N(x)$, $M_w(x)$, $M_\theta(x)$ and $\hat{Q}_\theta(x)$ are defined as

$$N(x)=\int_A\sigma_x(x,y)\mathrm{d}A \tag{16}$$

$$M_w(x)=\int_A\left[y-f(y)\right]\sigma_x(x,y)\mathrm{d}A \tag{17}$$

$$M_\theta(x)=\int_A f(y)\sigma_x(x,y)\mathrm{d}A \tag{18}$$

$$\hat{Q}_\theta(x)=\int_A f_{,y}(y)\tau_{xy}(x,y)\mathrm{d}A \tag{19}$$

with $A$ the beam's cross-section domain. It should also be noted that the positive directions of the internal forces are consistent with their corresponding generalized strains defined in Eqs. (5)-(8).

According to the definition of internal forces, $N(x)$ represents the axial force of the beam, $M_w(x)$, $M_\theta(x)$ and $\hat{Q}_\theta(x)$ are the internal forces conjugated with $\delta\kappa_w$, $\delta\kappa_\theta$ and $\delta\gamma_0$, respectively. The ratio of $M_w(x)$ to $M_\theta(x)$ is influenced by the definition of $f(y)$. The total bending moment of the beam consists of $M_w(x)$ and $M_\theta(x)$, namely

$$M(x)=\int_A y\sigma_x(x,y)\mathrm{d}A=M_w(x)+M_\theta(x), \tag{20}$$

and the total shear force of the beam can be obtained according to the following formula

$$Q(x)=M_{,x}(x)=M_{w,x}(x)+M_{\theta,x}(x) \tag{21}$$

By substituting the stress expressions (Eqs. (12) and (13)) and the strain expressions (Eqs. (3) and (4)) into Eqs. (16)-(19) and integrating the cross-section domain, the constitutive relation of the beam's cross-section can be obtained as follows

$$\boldsymbol{\sigma}(x)=\mathbf{D}_\mathrm{n}\boldsymbol{\varepsilon}(x) \tag{22}$$

$$\hat{Q}_\theta(x) = \hat{D}_s \gamma_0(x) \tag{23}$$

where $\hat{D}_s$ is referred to the approximate shear stiffness of the beam's cross-section, expressed as

$$\hat{D}_s = \int_A f_{,y}^2(y) G(y) \mathrm{d}A \tag{24}$$

In Eqs. (23) and (24), the hat is used to denote that $\hat{Q}_\theta(x)$ and $\hat{D}_s$ do not strictly satisfy the equilibrium relationship.

In Eq. (22), $\mathbf{D}_n$ represents the cross-section stiffness matrix related to normal stress, which can be obtained as

$$\mathbf{D}_n = \int_A E(y) \mathbf{t}^T(y) \mathbf{t}(y) \mathrm{d}A, \tag{25}$$

and $\boldsymbol{\sigma}(x)$ denotes the internal force vector associated with the normal stress, which is expressed as

$$\boldsymbol{\sigma}(x) = \{N(x) \quad M_w(x) \quad M_\theta(x)\}^T \tag{26}$$

For asymmetric FG material distributions, the non-diagonal components of the cross-sectional stiffness matrix $\mathbf{D}_n$ are non-zero, reflecting the coupling relationship between axial force and bending moments.

In accordance with Eq. (22), strain vector $\boldsymbol{\varepsilon}(x)$ can be expressed by using the internal force vector $\boldsymbol{\sigma}(x)$ as

$$\boldsymbol{\varepsilon}(x) = \mathbf{F}_n \boldsymbol{\sigma}(x) \tag{27}$$

where $\mathbf{F}_n$ is the cross-sectional flexibility matrix expressed as

$$\mathbf{F}_n = \mathbf{D}_n^{-1} = \begin{bmatrix} f_{11} & f_{12} & f_{13} \\ f_{12} & f_{22} & f_{23} \\ f_{13} & f_{23} & f_{33} \end{bmatrix} \tag{28}$$

2.4 Stress expressions

For FG beams where the material properties are intricately distributed along the thickness of the beam, the expression of transverse shear stress obtained by Eq. (13) cannot strictly satisfy the equilibrium relation. Therefore, it cannot reflect the true distribution of transverse shear stress. In order to obtain the rational distribution of transverse shear stress, the following differential equilibrium equation is used [9]

$$\frac{\partial \sigma_x(x,y)}{\partial x} + \frac{\partial \tau_{xy}(x,y)}{\partial y} = 0 \tag{29}$$

Firstly, based on Eqs. (10), (12) and (27), the axial normal stress can be expressed as

$$\sigma_x(x,y) = E(y) \mathbf{t}(y) \mathbf{F}_n \boldsymbol{\sigma}(x) \tag{30}$$

Then, by substituting Eq. (30) into Eq. (29) and integrating along the beam's thickness, the following relation can be derived

$$\tau_{xy} \mathrm{d}x = -\int_{-h/2}^{y} E(\xi) \mathbf{t}(\xi) \mathrm{d}\xi \mathbf{F}_n \mathrm{d}\boldsymbol{\sigma} \tag{31}$$

Further, the expression of transverse shear stress which satisfies the equilibrium relation can be obtained as

$$\tau_{xy} = \mathbf{S}(y) \boldsymbol{\sigma}_{,x} \tag{32}$$

where

$$\mathbf{S}(y) = -\int_{-h/2}^{y} E(\xi) \mathbf{t}(\xi) \mathrm{d}\xi \mathbf{F}_n = \left[ S_1(y) \quad S_2(y) \quad S_3(y) \right] \mathbf{F}_n \tag{33}$$

with

$$S_1(y) = -\int_{-h/2}^{y} E(\xi)\mathrm{d}\xi$$

$$S_2(y) = -\int_{-h/2}^{y} \left[\xi - f(\xi)\right] E(\xi)\mathrm{d}\xi \tag{34}$$

$$S_3(y) = -\int_{-h/2}^{y} f(\xi) E(\xi)\mathrm{d}\xi$$

Considering that the influence of axial force on shear force is relatively small, the impact of axial force is ignored, thereby obtaining the following simplified expression of the transverse shear stress as

$$\tau_{xy}(x,y) = \mathbf{S}(y)\boldsymbol{\tau}(x) \tag{35}$$

where

$$\boldsymbol{\tau}(x) = \left\{ M_{w,x}(x) \quad M_{\theta,x}(x) \right\}^{\mathrm{T}} \tag{36}$$

$$\mathbf{S}(y) = \left\{ S_w(y) \quad S_\theta(y) \right\} \tag{37}$$

$$S_w(y) = S_1(y) f_{12} + S_2(y) f_{22} + S_3(y) f_{32} \tag{38}$$

$$S_\theta(y) = S_1(y) f_{13} + S_2(y) f_{23} + S_3(y) f_{33} \tag{39}$$

2.5 Modified shear stiffness

The shear stiffness obtained from Eq. (24) may result in significant solution errors due to the failure to describe the true transverse shear stress distribution. Although the mixed higher-order shear beam element model proposed by Li et al. [56] improve the solution accuracy, it cannot ensure the continuity of internal forces between elements, which may result in abrupt changes in the predicted stresses along the beam axis. Different from the work of Li et al. [56], this paper derives the modified shear stiffness to involve the effect of rational transverse shear stress determined by Eq. (35). Then, the beam element model constructed based on the modified shear stiffness can maintain the continuity of internal forces between elements.

First of all, the energy function expressed by strains and stresses can be written as [56]

$$U\left(\sigma_x, \tau_{xy}, \varepsilon_x, \gamma_{xy}\right) = \int_0^L \int_A \left[ \sigma_x \varepsilon_x + \tau_{xy}\gamma_{xy} - \frac{1}{2}\frac{\sigma_x^2}{E(y)} - \frac{1}{2}\frac{\tau_{xy}^2}{G(y)} \right] \mathrm{d}A\mathrm{d}x \tag{40}$$

By introducing the internal force parameter vector $\boldsymbol{\beta}$, which will be specifically defined in **Sec. 3.1**, the fields of internal forces in Eq. (30) and Eq. (35) can be expressed by $\boldsymbol{\sigma}(\boldsymbol{\beta})$ and $\boldsymbol{\tau}(\boldsymbol{\beta})$. Meanwhile, the fields of generalized strains in Eqs. (10) and (4) can be expressed as $\boldsymbol{\varepsilon}(\mathbf{d})$ and $\gamma_0(\mathbf{d})$ with $\mathbf{d}$ the introduced displacement vector including the components of $u(x)$, $w(x)$ and $\theta(x)$. Then, by introducing Eqs. (30), (35), (10) and (4) and integrating across the cross-section domain, the energy function expressed by the two types of field quantities can be rewritten as

$$U(\boldsymbol{\beta},\mathbf{d}) = \int_0^L \left[ \boldsymbol{\sigma}^{\mathrm{T}}(\boldsymbol{\beta})\boldsymbol{\varepsilon}(\mathbf{d}) + \boldsymbol{\tau}^{\mathrm{T}}(\boldsymbol{\beta})\mathbf{f}_s\gamma_0(\mathbf{d}) - \frac{1}{2}\boldsymbol{\sigma}^{\mathrm{T}}(\boldsymbol{\beta})\mathbf{F}\boldsymbol{\sigma}(\boldsymbol{\beta}) - \frac{1}{2}\boldsymbol{\tau}^{\mathrm{T}}(\boldsymbol{\beta})\mathbf{f}_{ss}\boldsymbol{\tau}(\boldsymbol{\beta}) \right] \mathrm{d}x \tag{41}$$

where $\mathbf{f}_s$ and $\mathbf{f}_{ss}$ can be expressed as [56]

$$\mathbf{f}_s = \int_A \left[ \mathbf{S}^{\mathrm{T}}(y) f_{,y}(y) \right] \mathrm{d}A \tag{42}$$

$$\mathbf{f}_{ss} = \int_A \left[ \frac{\mathbf{S}^{\mathrm{T}}(y)\mathbf{S}(y)}{G(y)} \right] \mathrm{d}A \tag{43}$$

Assuming the existence of the shear force $Q_\theta\left(x\right)$ that corresponds to the shear deformation $\gamma_0\left(x\right)$ and satisfys the equilibrium relation, the work done by this shear force can be written as

$$U_s = \int_0^L Q_\theta\left(x\right)\gamma_0\left(x\right)\mathrm{d}x \tag{44}$$

It is considered that the energy related to shear deformation in Eq. (41) should be consistent with those in Eq. (44). Therefore, the following equation can be established

$$\int_0^L \left[\boldsymbol{\tau}^{\mathrm{T}}\left(\boldsymbol{\beta}\right)\mathbf{f}_s\gamma_0\left(\mathbf{d}\right)-\frac{1}{2}\boldsymbol{\tau}^{\mathrm{T}}\left(\boldsymbol{\beta}\right)\mathbf{f}_{ss}\boldsymbol{\tau}\left(\boldsymbol{\beta}\right)\right]\mathrm{d}x-\int_0^L Q_\theta\left(x\right)\gamma_0\left(x\right)\mathrm{d}x=0 \tag{45}$$

The integral domains of the two parts on the left side of the above equation are consistent. Therefore, the following equation established based on the differential segment of the beam can be considered as a sufficient condition to ensure that Eq. (45) holds.

$$\boldsymbol{\tau}^{\mathrm{T}}\mathbf{f}_s\gamma_0-\frac{1}{2}\boldsymbol{\tau}^{\mathrm{T}}\mathbf{f}_{ss}\boldsymbol{\tau}-Q_\theta\gamma_0=0 \tag{46}$$

Based on the variation principle that $\delta\left(\boldsymbol{\tau}^{\mathrm{T}}\mathbf{f}_s\gamma_0-\frac{1}{2}\boldsymbol{\tau}^{\mathrm{T}}\mathbf{f}_{ss}\boldsymbol{\tau}-Q_\theta\gamma_0\right)=0$, the following equation can be obtained

$$\delta\gamma_0\left(\mathbf{f}_s^{\mathrm{T}}\boldsymbol{\tau}-Q_\theta\right)+\delta\boldsymbol{\tau}^{\mathrm{T}}\left(\mathbf{f}_s\gamma_0-\mathbf{f}_{ss}\boldsymbol{\tau}\right)=0 \tag{47}$$

Considering that $\delta\boldsymbol{\tau}$ and $\delta\gamma_0$ are arbitrary variations, the following two sets of equations can be derived

$$\mathbf{f}_s\gamma_0-\mathbf{f}_{ss}\boldsymbol{\tau}=\mathbf{0} \tag{48}$$

$$\mathbf{f}_s^{\mathrm{T}}\boldsymbol{\tau}-Q_\theta=0 \tag{49}$$

From Eq. (48), we have $\boldsymbol{\tau}=\mathbf{f}_{ss}^{-1}\mathbf{f}_s\gamma_0$. Then, the following equation can be further obtained by substituting $\boldsymbol{\tau}=\mathbf{f}_{ss}^{-1}\mathbf{f}_s\gamma_0$ into Eq. (49)

$$Q_\theta=D_{\mathrm{s}}\gamma_0 \tag{50}$$

where $D_{\mathrm{s}}$ represents the modified shear stiffness, and it can be obtained by

$$D_{\mathrm{s}}=\mathbf{f}_s^{\mathrm{T}}\mathbf{f}_{ss}^{-1}\mathbf{f}_s \tag{51}$$

It is worth emphasizing that the modified shear stiffness is the main characteristic that distinguishes the modified HSBT from the traditional HSBT. In the above derivation process, the transverse shear stress is always expressed by Eq. (35), which is derived from the differential equilibrium equation (Eq. ((29))). Therefore, the modified shear stiffness ensures the equilibrium relations. Specifically, for the case of homogeneous material, the modified shear stiffness will be consistent with that expressed in Eq. (24).

## 2.6 Differential equilibrium equations of the beam

Based on the geometric relations indicated in Eqs. (5)-(8), the variation of the generalized strains can be expressed as

$$\begin{aligned}
\delta\varepsilon_0\left(x\right)&=\delta u_{,x}\left(x\right)\\
\delta\kappa_w\left(x\right)&=-\delta w_{,xx}\left(x\right)\\
\delta\kappa_\theta\left(x\right)&=-\delta\theta_{,x}\left(x\right)\\
\delta\gamma_0\left(x\right)&=\delta w_{,x}\left(x\right)-\delta\theta\left(x\right)
\end{aligned} \tag{52}$$

Therefore, the virtual of the beam strain energy $\delta U$ and the virtual work done by the external force $\delta W$ can be further expressed as

$$\begin{aligned}\delta U &= \int_L \left[ N(x)\delta\varepsilon_0(x) + M_w(x)\delta\kappa_w(x) + M_\theta(x)\delta\kappa_\theta(x) + Q_\theta(x)\delta\gamma_0(x) \right]\mathrm{d}x \\ &= \int_L \left\{ N\delta u_{,x} - M_w \delta w_{,xx} - M_\theta \delta\theta_{,x} + Q_\theta \delta w_{,x} - Q_\theta \delta\theta \right\}\mathrm{d}x\end{aligned} \tag{53}$$

$$\delta W = -\int_L q(x)\delta w \mathrm{d}x - P_{x1}\delta u(0) - P_{x2}\delta u(L) - P_{y1}\delta w(0) - P_{y2}\delta w(L) + M_1\delta\theta(0) + M_2\delta\theta(L) \tag{54}$$

where $P_{xi}, P_{yi}$ and $M_i\ (i=1,2)$ represent the external nodal loads at the starting and ending points, respectively. It is noteworthy that only the transverse distributed load $q(x)$ is considered in order to simplify the formulation. Based on the principle of virtual work that $\delta U + \delta W = 0$, the following equation can be obtained

$$\begin{aligned}&-\int_L \left\{ N_{,x} \right\}\delta u \mathrm{d}x - \int_L \left\{ M_{w,xx} + Q_{\theta,x} + q \right\}\delta w \mathrm{d}x + \int_L \left\{ M_{\theta,x} - Q_\theta \right\}\delta\theta \mathrm{d}x \\ &+\left[ N(L) - P_{x2} \right]\delta u(L) - \left[ N(0) + P_{x1} \right]\delta u(0) + \left[ M_{w,x}(L) + Q_\theta(L) - P_{y2} \right]\delta w(L) \\ &-\left[ M_{w,x}(0) + Q_\theta(0) + P_{y1} \right]\delta w(0) - M_w(L)\delta w_{,x}(L) + M_w(0)\delta w_{,x}(0) \\ &-\left[ M_\theta(L) - M_2 \right]\delta\theta(L) + \left[ M_\theta(0) + M_1 \right]\delta\theta(0) = 0\end{aligned} \tag{55}$$

Thus, the differential equilibrium equation of the higher-order shear beam can be written as

$$N_{,x}(x) = 0 \tag{56}$$

$$M_{w,xx}(x) + Q_{\theta,x}(x) + q(x) = 0 \tag{57}$$

$$M_{\theta,x}(x) - Q_\theta(x) = 0 \tag{58}$$

The corresponding boundary conditions are expressed as

(1) $x = 0$

$$\begin{aligned}\delta u(0) = 0 \quad &\text{or} \quad N(0) + P_{x1} = 0 \\ \delta w(0) = 0 \quad &\text{or} \quad M_{w,x}(0) + Q_\theta(0) + P_{y1} = 0 \\ \delta w_{,x}(0) = 0 \quad &\text{or} \quad M_w(0) = 0 \\ \delta\theta(0) = 0 \quad &\text{or} \quad M_\theta(0) + M_1 = 0\end{aligned} \tag{59}$$

(2) $x = L$

$$\begin{aligned}\delta u(L) = 0 \quad &\text{or} \quad N(L) - P_{x2} = 0 \\ \delta w(L) = 0 \quad &\text{or} \quad M_{w,x}(L) + Q_\theta(L) - P_{y2} = 0 \\ \delta w_{,x}(L) = 0 \quad &\text{or} \quad M_w(L) = 0 \\ \delta\theta(L) = 0 \quad &\text{or} \quad M_\theta(L) - M_2 = 0\end{aligned} \tag{60}$$

It can be observed that the total shear force is expressed as $M_{w,x}(x) + Q_\theta(x)$ in Eqs. (59) and (60), while $M_{w,x}(x) + M_{\theta,x}(x)$ in Eq. (21). In fact, $M_{\theta,x}(x) = Q_\theta(x)$ holds under the condition of stress equilibrium. This can be easily proven. By taking the first-order derivative of Eq. (18) with respect to *x* and introducing the relation of Eq. (29). the expression of $M_{\theta,x}(x)$ can be obtained as

$$M_{\theta,x}(x) = -f(y)\tau_{xy}(x,y)\Big|_{-h/2}^{h/2} + \int_A g(y)\tau_{xy}(x,y)\mathrm{d}A \tag{61}$$

By comparing Eq. (61) and Eq. (19), it can be easily know that, $M_{\theta,x}(x) = Q_\theta(x)$ holds if $-f(y)\tau_{xy}(x,y)\Big|_{-h/2}^{h/2} = 0$ is satisfied. Since that the value of transverse shear stress at the upper and lower boundaries is zero while considering the equilibrium condition, it is true for $-f(y)\tau_{xy}(x,y)\Big|_{-h/2}^{h/2} = 0$ and hence $M_{\theta,x}(x) = Q_\theta(x)$ holds. In other words, when the equilibrium condition is satisfied, the total shear force of the beam can be expressed as

$$Q(x) = M_{w,x}(x) + Q_\theta(x) \tag{62}$$

**3 Finite element implementation**

3.1 Fields of internal forces

Despite three differential equilibrium equations as shown in Eqs. (56)-(58) have been obtained, they are insufficient to derive solutions for the four field quantities, including $N(x)$, $M_w(x)$, $M_\theta(x)$ and $Q_\theta(x)$. Therefore, an additional equation is required to achieve the solution.

In this work, the additional equation is constructed based on the constitutive equation and the relationship between generalized strains and displacements. Based on the relations in Eqs. (27), (6) and (7), the following expressions of $w_{,xx}$ and $\theta_{,x}$ can be obtained

$$\begin{aligned} w_{,xx} &= -f_{12}N - f_{22}M_w - f_{23}M_\theta \\ \theta_{,x} &= -f_{13}N - f_{23}M_w - f_{33}M_\theta \end{aligned} \tag{63}$$

Meanwhile, the following equation can be obtained from Eqs. (50) and (8)

$$w_{,x} - \theta = \frac{1}{D_\mathrm{s}} Q_\theta \tag{64}$$

Then, by taking the derivative of Eq. (64) with respect to $x$ and using Eq. (63), the additional equation can be derived as

$$Q_{\theta,x}(x) = a_1 N(x) + a_2 M_w(x) + a_3 M_\theta(x) \tag{65}$$

where

$$a_1 = D_\mathrm{s}(f_{13} - f_{12}), a_2 = D_\mathrm{s}(f_{23} - f_{22}), a_3 = D_\mathrm{s}(f_{33} - f_{23}) \tag{66}$$

According to Eq. (56), $N(x)$ is constant along the beam axis and can be expressed as

$$N(x) = c_0 \tag{67}$$

where $c_0$ is the coefficient to be determined.

Considering the relations in Eqs. (20) and (62), $M_\theta(x)$ and $Q_\theta(x)$ can be expressed by $M(x)$ and $Q(x)$ as

$$M_\theta(x) = M(x) - M_w(x) \tag{68}$$

$$Q_\theta(x) = Q(x) - M_{w,x}(x) \tag{69}$$

By taking the first-order derivative of Eq. (69) with respent to $x$ and substituting it into Eq. (57), the following equation can be obtained

$$Q_{,x}(x) + q(x) = 0 \tag{70}$$

Therefore, the total shear force of the beam is

$$Q(x) = c_1 - \int_0^x q(x)\mathrm{d}x = c_1 - I_q(x) \tag{71}$$

where $I_q(x) = \int_0^x q(\xi)\mathrm{d}\xi$ and $c_1$ is the coefficient to be determined. Furthermore, considering Eq. (20) and Eq. (21), the total bending moment of the beam can be obtained as

$$M(x) = c_2 + \int_0^x Q(\xi)\mathrm{d}\xi = c_2 + \int_0^x \left(c_1 - \int_0^\xi q(\eta)\mathrm{d}\eta\right)\mathrm{d}\xi = c_2 + c_1 x - I_{qq}(x) \tag{72}$$

where $I_{qq}(x) = \int_0^x \int_0^\xi q(\eta)\mathrm{d}\eta\mathrm{d}\xi$ and $c_2$ is the coefficient to be determined.

Since that the expressions of $N(x)$, $Q(x)$ and $M(x)$ have been obtained, the expressions of $M_\theta(x)$ and $Q_\theta(x)$ can be further derived through Eqs. (68) and (69) as long as the expressions of $M_w(x)$ is determined. By using Eqs. (68) and (69), Eq. (65) can be rewritten as

$$M_{w,xx}(x)+(a_2-a_3)M_w(x)=-a_1N(x)-a_3M(x)+Q_{,x}(x) \tag{73}$$

Further, by substituting Eqs. (67), (71) and (72) into the above equation, the following differential equation can be obtained

$$M_{w,xx}(x)+gM_w(x)=a_3I_{qq}(x)-q(x)-a_1c_0-a_3c_1x-a_3c_2 \tag{74}$$

where

$$g=a_2-a_3 \tag{75}$$

Eq. (74) is a second-order linear ordinary differential equation, and the eigen equation corresponding to its homogeneous equation can be represented as

$$r^2+g=0 \tag{76}$$

In general, $g<0$ and hence the the solutions for the case with a pair of virtual roots will be mainly introduced in this paper. For the other two cases such as $g\ge 0$, the formulation can also be derived through a similar method, and they are no longer specifically provided due to space limitations.

For the case of $g<0$, the general solution of the homogeneous equation is

$$\bar{M}_w(x)=c_3e^{\lambda x}+c_4e^{-\lambda x} \tag{77}$$

where

$$\lambda=\sqrt{-g} \tag{78}$$

Further, the particular solution of Eq. (74) is preset as

$$M_w^*(x)=b_1I_{qq}(x)+b_2q(x)+b_3x+b_4 \tag{79}$$

The introduction of Eq. (79) into Eq. (74) can derive

$$(gb_1-a_3)I_{qq}(x)+(b_1+gb_2+1)q(x)+(gb_3+a_3c_1)x+(gb_4+a_1c_0+a_3c_2)=0 \tag{80}$$

To ensure the constancy of Eq. (80), $b_1\sim b_3$ are taken as

$$b_1=\frac{a_3}{g},b_2=-\frac{1}{g}\left(1+\frac{a_3}{g}\right),b_3=-\frac{a_3}{g}c_1,b_4=-\frac{a_1}{g}c_0-\frac{a_3}{g}c_2 \tag{81}$$

Hence, the closed-form solution of $M_w(x)$ is

$$M_w(x)=-c_0\left(\frac{a_1}{g}\right)-c_1\left(\frac{a_3x}{g}\right)-c_2\left(\frac{a_3}{g}\right)+c_3e^{\lambda x}+c_4e^{-\lambda x}+\frac{a_3}{g}I_{qq}(x)-\frac{1}{g}\left(1+\frac{a_3}{g}\right)q(x) \tag{82}$$

The expressions of $M_\theta(x)$ and $Q_\theta(x)$ can be obtained by substituting Eq. (82) into Eqs. (68) and (69). $c_0\sim c_4$ can be considered as the internal force parameters, which determine the fields of internal forces. For clarity, they can be expressed using the internal force parameter vector as

$$\boldsymbol{\beta}=\{c_0\quad c_1\quad c_2\quad c_3\quad c_4\}^{\mathrm{T}} \tag{83}$$

Then, the internal force fields can be expressed as

$$\begin{Bmatrix} N(x) \\ M_w(x) \\ M_\theta(x) \\ Q_\theta(x) \end{Bmatrix} = \frac{1}{g}\begin{bmatrix} g & 0 & 0 & 0 & 0 \\ -a_1 & -a_3 x & -a_3 & g e^{\lambda x} & g e^{-\lambda x} \\ a_1 & a_2 x & a_2 & -g e^{\lambda x} & -g e^{-\lambda x} \\ 0 & a_2 & 0 & -g\lambda e^{\lambda x} & g\lambda e^{-\lambda x} \end{bmatrix}\boldsymbol{\beta} + \frac{I_{qq}(x)}{g}\begin{Bmatrix} 0 \\ a_3 \\ -a_2 \\ 0 \end{Bmatrix} + \frac{I_q(x)}{g}\begin{Bmatrix} 0 \\ 0 \\ 0 \\ -a_2 \end{Bmatrix} + \frac{q(x)}{g^2}\begin{Bmatrix} 0 \\ -a_2 \\ a_2 \\ 0 \end{Bmatrix} + \frac{q_{,x}(x)}{g^2}\begin{Bmatrix} 0 \\ 0 \\ 0 \\ a_2 \end{Bmatrix} \tag{84}$$

Meanwhile, $\boldsymbol{\tau}(x) = \{M_{w,x}(x) \quad M_{\theta,x}(x)\}^{\mathrm{T}}$ in shear stress expression (Eq. (36)) can also be obtained as

$$\boldsymbol{\tau}(x) = \begin{Bmatrix} M_{w,x}(x) \\ M_{\theta,x}(x) \end{Bmatrix} = \frac{1}{g}\begin{bmatrix} 0 & -a_3 & 0 & g\lambda e^{\lambda x} & -g\lambda e^{-\lambda x} \\ 0 & a_2 & 0 & -g\lambda e^{\lambda x} & g\lambda e^{-\lambda x} \end{bmatrix}\boldsymbol{\beta} + \frac{I_q(x)}{g}\begin{Bmatrix} a_3 \\ -a_2 \end{Bmatrix} + \frac{q_{,x}(x)}{g^2}\begin{Bmatrix} -a_2 \\ a_2 \end{Bmatrix} \tag{85}$$

3.2 Fields of generalized displacements

Different from the tranditional beam finite elment models that the generalized displacements are defined as independent fields of unknown quantities, the fields of generalized displacements in the present beam model are determined by the internal force fields based on the constitutive relations and geometric equations. According to Eq. (84), the internal force fields related to axial normal stress and transverse shear stress are respectively expressed as

$$\boldsymbol{\sigma}(x) = \mathbf{N}_\sigma(x)\boldsymbol{\beta} + \mathbf{F}_\sigma(x) \tag{86}$$

$$Q_\theta(x) = \mathbf{N}_\tau(x)\boldsymbol{\beta} + \mathbf{F}_\tau(x) \tag{87}$$

where $\mathbf{N}_\sigma(x)$, $\mathbf{N}_\tau(x)$, $\mathbf{F}_\sigma(x)$ and $\mathbf{F}_\tau(x)$ are correspond to Eq. (84), and they are expressed as

$$\mathbf{N}_\sigma(x) = \frac{1}{g}\begin{bmatrix} 1 & 0 & 0 & 0 & 0 \\ -a_1 & -a_3 x & -a_3 & g e^{\lambda x} & g e^{-\lambda x} \\ a_1 & a_2 x & a_2 & -g e^{\lambda x} & -g e^{-\lambda x} \end{bmatrix} \tag{88}$$

$$\mathbf{N}_\tau(x) = \frac{1}{g}\begin{bmatrix} 0 & a_2 & 0 & -g\lambda e^{\lambda x} & g\lambda e^{-\lambda x} \end{bmatrix} \tag{89}$$

$$\mathbf{F}_\sigma(x) = \frac{I_{qq}(x)}{g}\begin{Bmatrix} 0 \\ a_3 \\ -a_2 \end{Bmatrix} + \frac{q(x)}{g^2}\begin{Bmatrix} 0 \\ -a_2 \\ a_2 \end{Bmatrix} \tag{90}$$

$$\mathbf{F}_\tau(x) = -\frac{a_2 I_q(x)}{g} + \frac{a_2 q_{,x}(x)}{g^2} \tag{91}$$

Then, the expressions for each generalized displacement can be given as follows.

(1) Axial displacement

By integrating $u_{,x}(x)$, the expression of axial displacement can be expressed as

$$u(x) = u^a + \int_0^x u_{,x}(\xi)\,\mathrm{d}\xi \tag{92}$$

where $u^a$ represents the axial displacement of the beam's centre line at the starting node. By introducing Eqs. (5), (27) and (86), Eq. (92) can be further expressed as

$$u(x) = u^a + \mathbf{N}_u(x)\boldsymbol{\beta} + \mathbf{U}_u(x) \tag{93}$$

where

$$\mathbf{N}_u(x) = \mathbf{T}_u \mathbf{F}_{\mathrm{n}} \int_0^x \mathbf{N}_\sigma(\xi)\,\mathrm{d}\xi \tag{94}$$

$$\mathbf{U}_u(x) = \mathbf{T}_u \mathbf{F}_{\mathrm{n}} \int_0^x \mathbf{F}_\sigma(\xi)\,\mathrm{d}\xi \tag{95}$$

$$\mathbf{T}_u = \begin{bmatrix} 1 & 0 & 0 \end{bmatrix} \tag{96}$$

(2) Rotation of the cross-section

By integrating $\theta_{,x}(x)$, the expression of rotation can be expressed as

$$\theta(x) = \theta^a + \int_0^x \theta_{,x}(\xi)\mathrm{d}\xi \tag{97}$$

where $\theta^a$ is the rotatin of the starting node. In accordance of Eqs. (7), (27) and (86), Eq. (97) can be expressed as

$$\theta(x) = \theta^a + \mathbf{N}_\theta(x)\boldsymbol{\beta} + \mathbf{U}_\theta(x) \tag{98}$$

where

$$\mathbf{N}_\theta(x) = \mathbf{T}_\theta \mathbf{F}_\mathrm{n} \int_0^x \mathbf{N}_\sigma(\xi)\mathrm{d}\xi \tag{99}$$

$$\mathbf{U}_\theta(x) = \mathbf{T}_\theta \mathbf{F}_\mathrm{n} \int_0^x \mathbf{F}_\sigma(\xi)\mathrm{d}\xi \tag{100}$$

$$\mathbf{T}_\theta = \begin{bmatrix} 0 & 0 & -1 \end{bmatrix} \tag{101}$$

(3) First-order derivative of transverse displacement

By integrating $w_{,xx}(x)$, the expression of $w_{,x}(x)$ can be expressed as

$$w_{,x}(x) = w_{,x}^a + \int_0^x w_{,xx}(\xi)\mathrm{d}\xi \tag{102}$$

where $w_{,x}^a$ is the first-order derivative of transverse displacement at the starting node. Based on Eqs. (6), (27) and (86), Eq. (102) can be further expressed as

$$w_{,x}(x) = w_{,x}^a + \mathbf{N}_{ww}(x)\boldsymbol{\beta} + \mathbf{U}_{ww}(x) \tag{103}$$

where

$$\mathbf{N}_{ww}(x) = \mathbf{T}_w \mathbf{F}_\mathrm{n} \int_0^x \mathbf{N}_\sigma(\xi)\mathrm{d}\xi \tag{104}$$

$$\mathbf{U}_{ww}(x) = \mathbf{T}_w \mathbf{F}_\mathrm{n} \int_0^x \mathbf{F}_\sigma(\xi)\mathrm{d}\xi \tag{105}$$

$$\mathbf{T}_w = \begin{bmatrix} 0 & -1 & 0 \end{bmatrix} \tag{106}$$

(4) Transverse displacement

It is noteworthy that Eqs. (6) and (8) provide two different ways to derive the transverse displacement. The transverse displacement derived from Eq. (6) is related to the bending deformation, while the transverse displacement derived from Eq. (8) is related to the shear deformation. In other words, there are two different expressions for the transverse displacement.

Firstly, by integrating $w_{,x}(x)$, the expression of the transverse displacement can be expressed as

$$w(x) = w^a + \int_0^x w_{,x}(\xi)\mathrm{d}\xi \tag{107}$$

where $w^a$ is the transverse displacement of the centre line at starting node.

Subsequently, by substituting Eq. (103) into Eq. (107), the transverse displacement related to bending deformation can be expressed as

$$w(x) = w^a + xw_{,x}^a + \mathbf{N}_w(x)\boldsymbol{\beta} + \mathbf{U}_w(x) \tag{108}$$

where

$$\mathbf{N}_w(x)=\mathbf{T}_w\mathbf{F}_n\int_0^x\left[\int_0^\xi\mathbf{N}_\sigma(\eta)\mathrm{d}\eta\right]\mathrm{d}\xi \tag{109}$$

$$\mathbf{U}_w(x)=\mathbf{T}_w\mathbf{F}_n\int_0^x\left[\int_0^\xi\mathbf{F}_\sigma(\eta)\mathrm{d}\eta\right]\mathrm{d}\xi \tag{110}$$

Finally, by substituting Eqs. (8), (98) and (50) into Eq. (107), the transverse displacement related to shear deformation, which is represented as $w_s(x)$ for differentiation, can be expressed as

$$w_s(x)=w^a+x\theta^a+\mathbf{N}_{sw}(x)\boldsymbol{\beta}+\mathbf{U}_{sw}(x) \tag{111}$$

where

$$\mathbf{N}_{sw}(x)=\int_0^x\mathbf{N}_\theta(\xi)\mathrm{d}\xi+D_s^{-1}\int_0^x\mathbf{N}_\tau(\xi)\mathrm{d}\xi \tag{112}$$

$$\mathbf{U}_{sw}(x)=\mathbf{T}_\theta\mathbf{F}_n\int_0^x\left[\int_0^\xi\mathbf{F}_\sigma(\eta)\mathrm{d}\eta\right]\mathrm{d}\xi+D_s^{-1}\int_0^x\mathbf{F}_\tau(\xi)\mathrm{d}\xi \tag{113}$$

It can be obsversed that two expressions of the transverse displacement derived from constitutive relations and geometric equations (Eqs. (108) and (111)) are different. For a beam element model, these two transverse displacements should remain consistent at both the beam's starting and ending nodes.

### 3.3 Element equations

Based on the analytical expressions of internal forces, the equation system of a higher-order beam element can be constructed through the equilibrium conditions at the element boundaries (the starting and ending nodes) and the compatibility condition within the element.

The proposed two-node beam element has 8 displacement unknowns and 5 internal force parameters ($c_0 \sim c_4$). The 8 displacement unknowns are $u^a, w^a, w^a_{,x}, \theta^a, u^b, w^b, w^b_{,x}, \theta^b$, corresponding to the 4 displacement Degrees of Freedom (DoFs) at each of the two nodes (denoted by *a* and *b*, respectively). In other words, there are a total of 13 unknowns to be solved. Therefore, 13 equations should be set up to establish the equation system for each element.

(1) Equations of boundary condition

Since that the analytical expressions of internal force fields have been obtained, the equations for the equilibrium relations at the two nodes can be set up based on the boundary conditions listed in Eqs. (59) and (60). For simplity, a vector to express the internal force fields corresponding to the components listed in Eqs. (59) and (60) is defined as

$$\mathbf{S}(x)=\mathbf{P}(x)\boldsymbol{\beta}+\mathbf{F}(x) \tag{114}$$

where

$$\mathbf{S}(x)=\left\{N(x)\quad Q(x)\quad M_w(x)\quad M_\theta(x)\right\}^{\mathrm{T}} \tag{115}$$

$$\mathbf{P}(x)=\frac{1}{g}\begin{bmatrix} g & 0 & 0 & 0 & 0 \\ 0 & g & 0 & 0 & 0 \\ -a_1 & -a_3x & -a_3 & ge^{\lambda x} & ge^{-\lambda x} \\ a_1 & a_2x & a_2 & -ge^{\lambda x} & -ge^{-\lambda x} \end{bmatrix} \tag{116}$$

$$\mathbf{F}(x)=\frac{I_{qq}(x)}{g}\begin{Bmatrix} 0 \\ a_3 \\ -a_2 \\ 0 \end{Bmatrix}+\frac{I_q(x)}{g}\begin{Bmatrix} 0 \\ 0 \\ 0 \\ -a_2 \end{Bmatrix}+\frac{q(x)}{g^2}\begin{Bmatrix} 0 \\ -a_2 \\ a_2 \\ 0 \end{Bmatrix}+\frac{q_{,x}(x)}{g^2}\begin{Bmatrix} 0 \\ 0 \\ 0 \\ a_2 \end{Bmatrix} \tag{117}$$

Then, the 8 equilibrium equations corresponding to the starting node $a$ and the ending node $b$ can be repectively expressed as

$$\mathbf{S}_a+\mathbf{S}(0)=\mathbf{0}\Rightarrow-\mathbf{P}(0)\boldsymbol{\beta}=\mathbf{S}_a+\mathbf{F}(0) \tag{118}$$

$$\mathbf{S}_b-\mathbf{S}(L)=\mathbf{0}\Rightarrow\mathbf{P}(L)\boldsymbol{\beta}=\mathbf{S}_b-\mathbf{F}(L) \tag{119}$$

where $\mathbf{S}_a$ and $\mathbf{S}_b$ are the external forces applying on the starting node $a$ and the ending node $b$ of the beam element, $L$ is the length of the beam element.

(2) Equations of compatibility condition

Considering the consistency of the generalized displacements at the ending node between the nodal displacements and corresponding values obtained from the generalized displacement fields, the following equations of deformation compatibility can be established.

$$\begin{Bmatrix} u(L) \\ w(L) \\ w_{,x}(L) \\ \theta(L) \\ w_s(L) \end{Bmatrix}-\begin{Bmatrix} u^b \\ w^b \\ w_{,x}^b \\ \theta^b \\ w^b \end{Bmatrix}=\begin{Bmatrix} 0 \\ 0 \\ 0 \\ 0 \\ 0 \end{Bmatrix} \tag{120}$$

where $u^b, w^b, w_{,x}^b, \theta^b$ are the generalized displacement components at the ending node. It is noteworthy that, even though the state of a node is described by 4 generalized displacement components ($u^b, w^b, w_{,x}^b, \theta^b$), 5 equations can be established because the consistency of transverse displacement at the ending node should holds for both fields described by Eqs. (108) and (111). By substituting Eqs. (93), (98), (103), (108) and (111) with $x=L$, Eq. (120) can be rewritten as

$$\mathbf{N}_a\boldsymbol{\varphi}^a+\mathbf{N}_b\boldsymbol{\varphi}^b+\mathbf{N}_\beta(L)\boldsymbol{\beta}+\mathbf{U}(L)=\mathbf{0} \tag{121}$$

where

$$\begin{aligned} \boldsymbol{\varphi}^a&=\begin{Bmatrix} u^a & w^a & w_{,x}^a & \theta^a \end{Bmatrix}^{\mathrm{T}} \\ \boldsymbol{\varphi}^b&=\begin{Bmatrix} u^b & w^b & w_{,x}^b & \theta^b \end{Bmatrix}^{\mathrm{T}} \end{aligned} \tag{122}$$

$$\mathbf{N}_a=\begin{bmatrix} 1 & 0 & 0 & 0 \\ 0 & 1 & L & 0 \\ 0 & 0 & 1 & 0 \\ 0 & 0 & 0 & 1 \\ 0 & 1 & 0 & L \end{bmatrix},\ \mathbf{N}_b=\begin{bmatrix} -1 & 0 & 0 & 0 \\ 0 & -1 & 0 & 0 \\ 0 & 0 & -1 & 0 \\ 0 & 0 & 0 & -1 \\ 0 & -1 & 0 & 0 \end{bmatrix} \tag{123}$$

$$\mathbf{N}_\beta(L)=\begin{bmatrix} \mathbf{N}_u(L) \\ \mathbf{N}_w(L) \\ \mathbf{N}_{ww}(L) \\ \mathbf{N}_\theta(L) \\ \mathbf{N}_{sw}(L) \end{bmatrix},\ \mathbf{U}(L)=\begin{bmatrix} \mathbf{U}_u(L) \\ \mathbf{U}_w(L) \\ \mathbf{U}_{ww}(L) \\ \mathbf{U}_\theta(L) \\ \mathbf{U}_{sw}(L) \end{bmatrix} \tag{124}$$

(3) Equation system of the element

By integrating the equations of boundary condition at both nodes with the equations of compatibility condition, the following equation system can be obtained for each beam element

$$\begin{bmatrix} \mathbf{0}_{4\times4} & \mathbf{0}_{4\times4} & -\mathbf{P}(0) \\ \mathbf{0}_{4\times4} & \mathbf{0}_{4\times4} & \mathbf{P}(L) \\ \mathbf{N}_a & \mathbf{N}_b & \mathbf{N}_\beta(L) \end{bmatrix} \begin{Bmatrix} \boldsymbol{\varphi}^a \\ \boldsymbol{\varphi}^b \\ \boldsymbol{\beta} \end{Bmatrix} = \begin{Bmatrix} \mathbf{S}_a \\ \mathbf{S}_b \\ \mathbf{0} \end{Bmatrix} + \begin{Bmatrix} \mathbf{F}(0) \\ -\mathbf{F}(L) \\ -\mathbf{U}(L) \end{Bmatrix} \tag{125}$$

3.4 Equation system of the structure

Similar to the assembly method described in Ref. [69], the equation system for a structure with mutilple elements can be constructed by aggregating the equations of all elements. In this section, a cantilever beam composed of two higher-order beam elements is taken as an example to illustrate the construction of the DoFs and structural equations, as shown in **Fig. 3**.

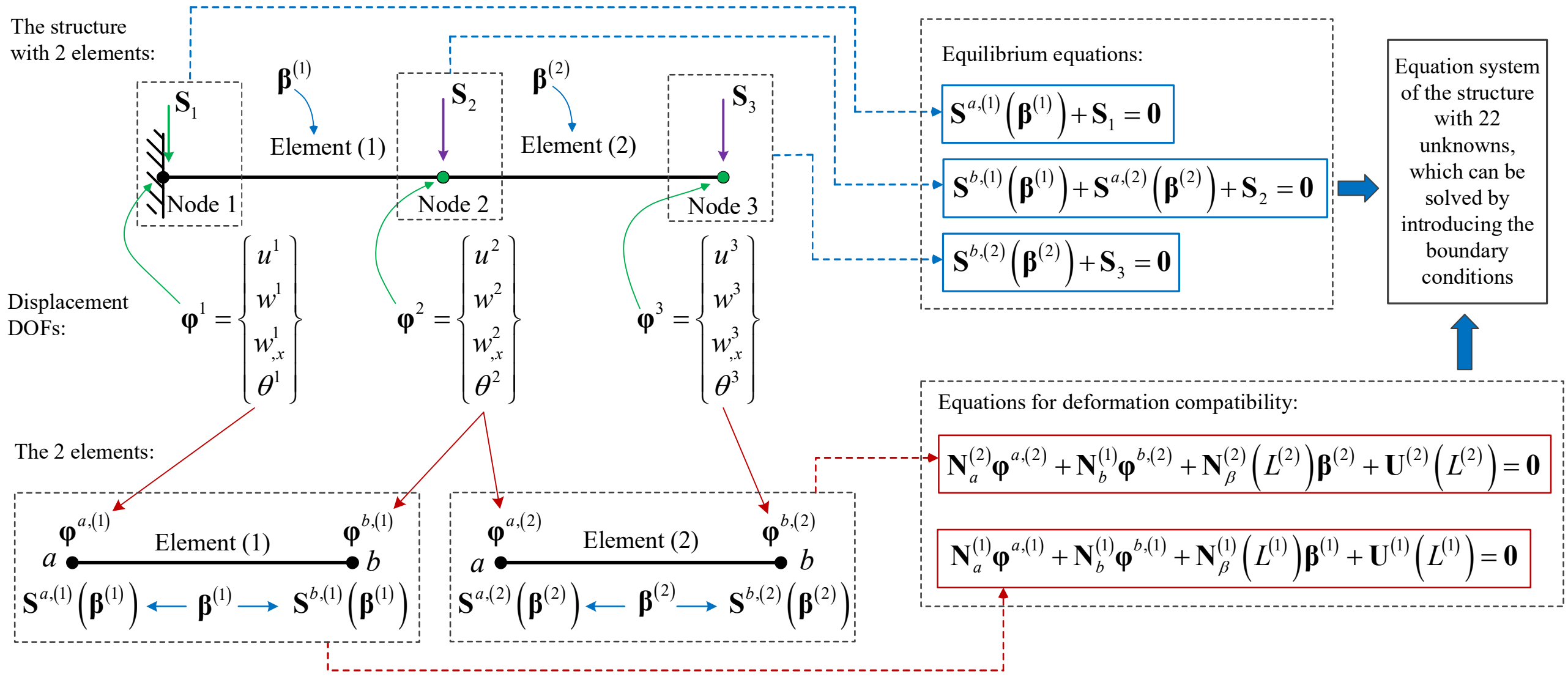

**Fig. 3**. Degrees of freedom and equation system of the structure.

In **Fig. 3**, the structure with three nodes (nodes 1, 2 and 3) has 12 displacement DoFs, which are expressed by $\boldsymbol{\varphi}^1$, $\boldsymbol{\varphi}^2$ and $\boldsymbol{\varphi}^3$ for the corresponding nodes. The displacement DoFs of element ($i$) ($i$ = 1, 2) are denoted as $\boldsymbol{\varphi}^{a,(i)}$ and $\boldsymbol{\varphi}^{b,(i)}$ for the two ends of the element. Generally, for two connected elements, the displacement components at their intersection are consistent, and thus are designated as the same unknowns in the structural system. Therefore, the relationships of element displacement DoFs and structural displacement DoFs can be expressed as

$$\begin{aligned} &\boldsymbol{\varphi}^{a,(1)} = \boldsymbol{\varphi}^1, \\ &\boldsymbol{\varphi}^{b,(1)} = \boldsymbol{\varphi}^{a,(2)} = \boldsymbol{\varphi}^2 \\ &\boldsymbol{\varphi}^{b,(2)} = \boldsymbol{\varphi}^3 \end{aligned} \tag{126}$$

In addition, 10 internal force parameters (5 for each element) are included in the structure and they are expressed by $\boldsymbol{\beta}^{(1)}$ and $\boldsymbol{\beta}^{(2)}$ for the corresponding elements. Then, the internal forces at the two ends ($a$ and $b$) for element ($i$) ($i$ = 1, 2) can be determined by $\boldsymbol{\beta}^{(i)}$ and expressed as $\mathbf{S}^{a,(i)}(\boldsymbol{\beta}^{(i)})$ and $\mathbf{S}^{b,(i)}(\boldsymbol{\beta}^{(i)})$. In **Fig. 3**, $\mathbf{S}_1$, $\mathbf{S}_2$ and $\mathbf{S}_3$ represent the external forces applying on nodes 1, 2 and 3, respectively.

The equation system of the structure can be constructed based on the equailibirum relationships at each node and the deformation compatibility within each element. The specific equations are shown in **Fig. 3** for the cantilever beam composed of two elements. Then, the total number of equations for the structure is $4n_{\text{node}} + 5n_{\text{elt}}$, where $n_{\text{node}}$ and $n_{\text{elt}}$ represent the total number of nodes and the number of elements, respectively. It should be pointed out that the total number of unknowns is also $4n_{\text{node}} + 5n_{\text{elt}}$ (without introducing the boundary conditions), which matches the number of equations. The equation system of the structure can be rewritten as a linear equation system represented in matrix form, where each column in the coefficient matrix should correspond to an unknown variable in the structural system, and each row in the coefficient matrix aligns with the equation constituting the structural system. Finally, the equations can be solved after introducing the boundary conditions. For the cantilever beam shown in **Fig. 3**, the equilibrium equations for node 1 can be removed and the displacements at node 1 can be considered fixed. Then, the size of equation system to be solved is 18.

3.5 Numerical integration

In the implementation of the element equations, numerical integration is used to generate all components in $\mathbf{N}_\beta(L)$ and $\mathbf{U}(L)$. Specifically, $\mathbf{N}_u(L), \mathbf{N}_\theta(L), \mathbf{N}_{ww}(L), \mathbf{N}_{sw}(L), \mathbf{U}_u(L), \mathbf{U}_\theta(L), \mathbf{U}_{ww}(L)$ and $\mathbf{U}_{ww}(L)$ can be generated throuhg a single integral, while a double integral is required to obtain the values of $\mathbf{N}_w(L), \mathbf{U}_w(L)$ and $\mathbf{U}_{sw}(L)$. In the double integral, a set of additional local (or internal) integrals needs to be evaluated numericaly. Therefore, a global-local integration scheme is used in this study, and the settings of integral points and the integral weights are shown in **Fig. 4**.

● Global integral points for double integral (Equidistant distribution) $\quad w_1 = w_m = \dfrac{0.5L}{m-1} \quad w_i = \dfrac{L}{m-1} \quad (i = 2,3,\ldots,m-1)$

○ Local integral points for single integral

$x_i^1, x_i^2, x_i^3 \ (i = 2,3,\ldots,m)$ are local Gauss integral points in $[x_{i-1}, x_i]$

$w_i^1, w_i^2, w_i^3 \ (i = 2,3,\ldots,m)$ are local Gauss integral weights in $[x_{i-1}, x_i]$

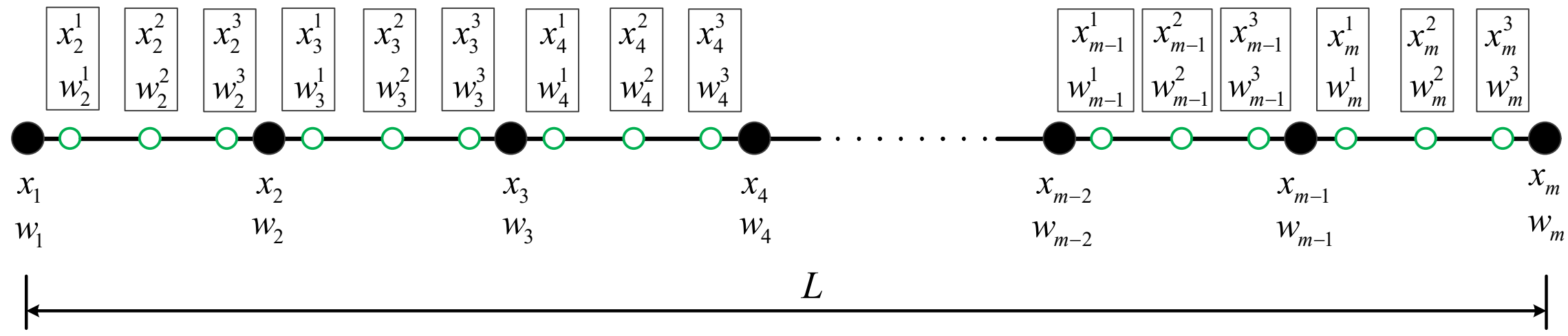

**Fig. 4**. The settings of integral points and the integral weights.

As shown in **Fig. 4**, $x_i\ (i = 1,2,\ldots,m)$ and $w_i\ (i = 1,2,\ldots,m)$ represent the global integral points and integral weights for the double integral, and the equidistant distribution including the boundary nodes is used for setting of $x_i\ (i = 1,2,\ldots,m)$. Then, the integral weights are set as $w_i = L/(m-1)(i = 2,2,\ldots,m-1)$ and $w_1 = w_m = 0.5L/(m-1)$. In addition, for each interval of $[x_{i-1}, x_i](i = 2,3,\ldots,m)$, the 3-point local Gaussian integration is adopted and the integral points and integral weights are denoted as $x_i^j\ (j = 1,2,3)$ and $w_i^j\ (j = 1,2,3)$. Generally, $x_i^j\ (j = 1,2,3)$ and $w_i^j\ (j = 1,2,3)$ can be directly used for the single integral. In the implementation of the double integral, the function value at $x_i\ (i =,2,\ldots,m)$ can be obtained through a step-by-step computation of local integrals using $x_i^j\ (j = 1,2,3)$ and $w_i^j\left(j=1,2,3\right)$, and subsequently the double integral can be achieved using $w_i\left(i=1,2,\ldots,m\right)$ for each global integral points. In this study, the total number of global integral points is set to $m=200$ to produce accurate results.

The above global-local numerical integration scheme is also used to obtain accurate values of $\mathbf{f}_s$ and $\mathbf{f}_{ss}$ in Eqs. (42) and (43). For FG material models with multiple layers, the global-local numerical integration scheme is used for each layer.

**4 Numerical examples**

In this section, two numerical examples are conducted to demonstrate the accuracy and effectivity of the proposed beam element. Several finite element models used in the investigation are introduced as follows:

(1) DEB – the **D**isplacement-based beam element based on **E**uler **B**eam theory,

(2) DFS – the **D**isplacement-based beam element based on **F**irst-order **S**hear deformation theory,

(3) DTS – the **D**isplacement-based beam element based on traditional **T**hird-order **S**hear deformation theory,

(4) MTS – the **Mi**xed beam element based on **T**hird-order **S**hear deformation theory [56],

(5) PFTS – the proposed beam element based on **P**redefined **F**orce fields and modified **T**hird-order **S**hear deformation theory,

(6) PFTS-T – the beam element based on **P**redefined **F**orce fields and **T**hird-order **S**hear deformation theory with **T**randitional shear stiffness,

(7) Q4 – the **4**-node Planar **Q**uadrilateral Element.

The constitutive relations (Eq. (12) or Eqs. (12) and (13)) are employed in the beam elements, including DEB, DFS, DTS, MTS, PFTS-T and PFTS. PFTS and PFTS-T are implemented based on the formulation provided in this paper, with different shear stiffness. The conventinal shear stiffness $\widehat{D}_\mathrm{s}$ is used in PFTS-T, while the modified shear stiffness $D_\mathrm{s}$ is adopted in PFTS. In other words, PFTS-T is a degraded version of PFTS that does not consider reasonable shear stress distribution.

DEB, DFS and DTS are beam elements establised based on variation principle of strain energy. In DEB, the displacement fields are defined as [3]

$$\begin{aligned} u_x\left(x,y\right) &= u\left(x\right)-y\frac{\mathrm{d}w\left(x\right)}{\mathrm{d}x} \\ u_y\left(x,y\right) &= w\left(x\right) \end{aligned} \tag{127}$$

The generalized displacement fields $u\left(x\right)$ and $w\left(x\right)$ are considered as the unknown fields. In the implementation of beam element, linear interpolation and cubic Hermite interpolation are used to discrete the axial displacement $u\left(x\right)$ and transverse displacement $w\left(x\right)$, respectively. In DFS, the displacement fields are expressed as [9]

$$\begin{aligned} u_x\left(x,y\right) &= u\left(x\right)-y\theta\left(x\right) \\ u_y\left(x,y\right) &= w\left(x\right) \end{aligned} \tag{128}$$

The unknown fields in DFS include axial displacement $u\left(x\right)$, transverse displacement $w\left(x\right)$ and rotation $\theta\left(x\right)$, and they are discreted by using linear interpolation. Particularly, the shear correction factor required by DFS is set to 5/6, since that the beam's cross-section is rectangular. DTS has the same definition of displacement fields as given in Eq. (1). Different from the PFTS and PFTS-T, DTS is derived from the variational principle of strain energy, and its unknown fields include axial displacement $u(x)$, transverse displacement $w(x)$ and rotation $\theta(x)$. For discretization, linear interpolation is used for $u(x)$ and $\theta(x)$, while cubic Hermite interpolation is employed for $w(x)$.

In MTS, generalized displacements and internal forces are considered as two types of independent unknown fields, and the beam finite element is constructed according to the mixed variational principle. Especially, the expression of transverse shear stress is derived from Eq. (29), and hence the accurate distribution of transverse shear stress can be obtained. For generalized displacements, the interpolation method of MTS is consistent with those in DTS. In addition, the internal force fields in the beam element are described using polynomials. For details of MTS, readers are referred to Ref. [56].

Different from the beam element mentioned above, Q4 can fully accommodate various in-plane deformations. Generally, the accurate displacement and stress results can be obtained by Q4 with sufficient refinement of meshes. Therefore, in this investigation, the displacement and stress solutions obtained using Q4 on a sufficiently refined mesh are employed as the references for evaluating other beam element models.

The element models and corresponding solution algorithms provided in this section have been programed in MATLAB and run on a computer having an Intel® CoreTM i7-8700 processor and a CPU at 3.2GHz with 64GB of RAM.

4.1 FG material models

Three different types of FG material models with a mixture form of ceramic and metal materials are considered: isotropic FG model (Type A), sandwich model with FG faces and homogeneous core (Type B), and sandwich model with FG core and homogeneous faces (Type C). For each FG material model, the Young's modulus along the thickness, $E(y)$, is given in the following form

$$E(y)=E_m+(E_c-E_m)V_c(y) \tag{129}$$

where $E_m$ and $E_c$ are the Young's modulus of the ceramic material and metal material, respectively, and $V_c(y)$ is the volume fraction of ceramic material, which can be determined as follows for the three FG material models:

(a) Type A: isotropic FG model

$$V_c(y)=\left(\frac{y-h_0}{h_1-h_0}\right)^p \qquad for\ \ y\in[h_0,h_1] \tag{130}$$

(b) Type B: sandwich model with FG faces and homogeneous core

$$V_c(y)=\begin{cases}\left[(y-h_0)/(h_1-h_0)\right]^p & for\ \ y\in[h_0,h_1]\\ 1 & for\ \ y\in[h_1,h_2]\\ \left[(y-h_3)/(h_2-h_3)\right]^p & for\ \ y\in[h_2,h_3]\end{cases} \tag{131}$$

(c) Type C: sandwich model with FG core and homogeneous faces

$$V_c(y)=\begin{cases}0 & for\ \ y\in[h_0,h_1]\\ \left[(y-h_1)/(h_2-h_1)\right]^p & for\ \ y\in[h_1,h_2]\\ 1 & for\ \ y\in[h_2,h_3]\end{cases} \tag{132}$$

where $p$ is the power-law index, $h_0,h_1,h_2,h_3$ are characteristic positions related to material distribution, including the junction position of adjacent material layers and the boundary position of beam's thickness, as shown in **Fig. 5**.

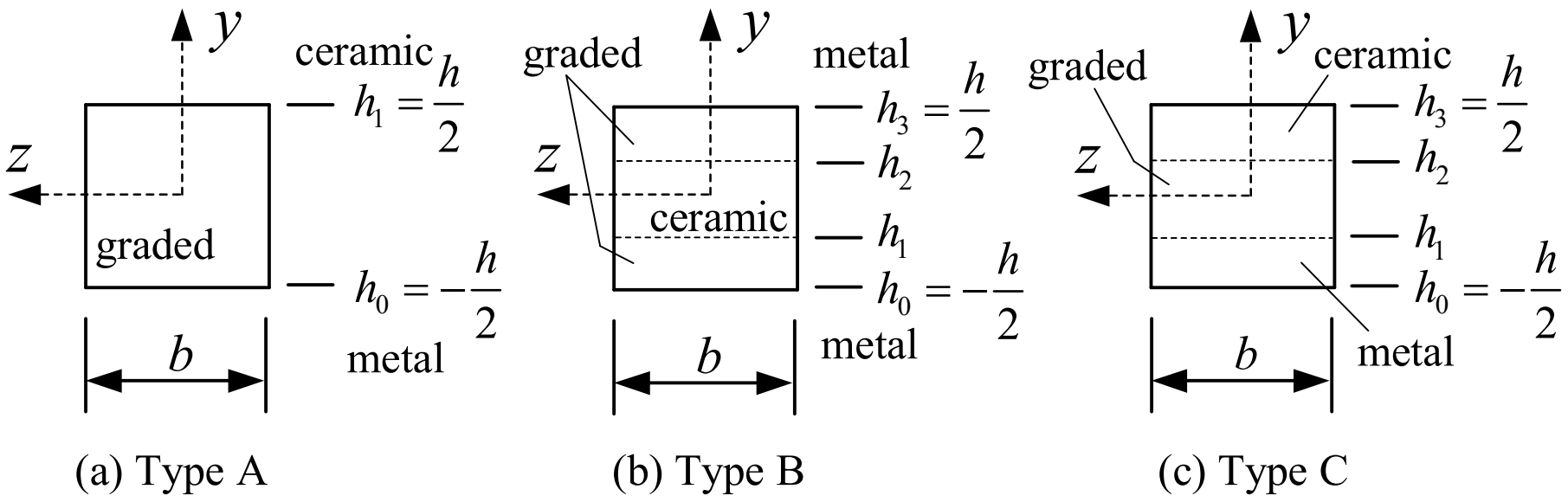

**Fig. 5**. The characteristic positions related to material distribution.

The FG material properties are set to be [56]: Aluminum ( $\text{Al}: E_m = 70000\,\text{N/mm}^2$ ) and Alumina ($Al_2O_3$: $\text{Al}: E_c = 380000\,\text{N/mm}^2$ ). The Poisson's ratio of material is set to $\nu = 0.3$ . For the beams with Type A material distribution, the characteristic positions are set to $h_0 = -100\text{mm}$ and $h_1 = 100\text{mm}$ , while for the beams with Type B and Type C material distributions, the characteristic positions are set to $h_0 = -100\text{mm}, h_1 = -40\text{mm}, h_2 = 40\text{mm}$ and $h_3 = 100\text{mm}$ . The power-law index $p$ is set to 0, 0.5, 1.0, 5.0, 10.0, respectively. For the given cross-section and material parameters, the values of $g$ obtained by Eqs. (75) and (66) are presented in **Table 1**. It is shown that under the five settings of $p$, the values of $g$ for the three types of FG material models are all less than zero. In other words, Eq. (76) has two imaginary roots, and the internal force fields presented in Eq. (84) are appropriate for this study.

**Table 1** Value of $g$ in Eq. (75) for three types of FG materials

| $p$ | Type A | Type B | Type C |
| --- | --- | --- | --- |
| 0.0 | −0.0081 | −0.0081 | −0.0096 |
| 0.5 | −0.0087 | −0.0120 | −0.0086 |
| 1.0 | −0.0081 | −0.0149 | −0.0076 |
| 5.0 | −0.0056 | −0.0172 | −0.0054 |
| 10.0 | −0.0055 | −0.0168 | −0.0050 |

4.2 FG cantilever subjected to a vertical load

This section examines the cantilever beam (Clamped-Free, C-F) depicted in **Fig. 6**, characterized by a length of 1000mm and a concentrated load applied vertically at the beam's free end.

A convergence test is performed on various beam element models for the beams with Type B and Type C material models under the condition of $p$ = 5.0, with the outcomes presented in **Table 2** and **Table 3**. The convergence test reveals that the beam elements formulated in this paper (inclusive of PFTS and PFTS-T) can attain convergence with a single element. This suggests that the element models, which consider the internal forces as the unknown fields, can effectively circumvent discretization errors. In contrast to PFTS and PFTS-T, the other beam elements, including DEB, DFS, DTS and MTS, necessitate a progressive refinement of mesh to approach the converged solutions, especially for cases with asymmetric material distribution. Furthermore, the solution efficiency of all beam elements has been investigated. For Type B and Type C material models, the number of DoFs as well as the corresponding relative computational time required for obtaining the converged solutions are listed in **Table 4** and **Table 5**, respectively, where $T_{\text{DTS}}$ represents the computational time of DTS and $T_{\text{beam}}$ refers to the computational time of the other beam element models. It is demonstrated that, due to the requirement of only one element to obtain the converged solutions, the proposed beam element model has significant advantages in computational efficiency compared to other beam element models.

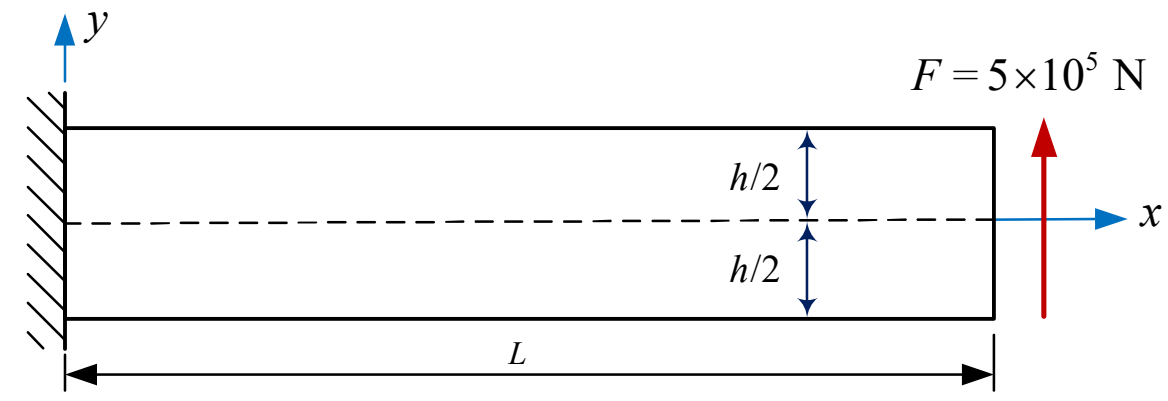

**Fig. 6**. Geometry of the cantilever.

**Table 2** Convergence of the tip displacement (mm) (C-F, Type B with $p$ = 5.0).

| Number of elements | DEB | DFS | DTS | MTS | PFTS-T | PFTS |
|---|---|---|---|---|---|---|
| 1 | 44.527 | 2.6514 | 34.044 | 45.132 | 45.088 | 45.102 |
| 2 | 44.527 | 9.0191 | 42.391 | 45.132 | 45.088 | 45.102 |
| 4 | | 22.571 | 44.468 | | | |
| 8 | | 36.151 | 44.966 | | | |
| 16 | | 42.552 | 45.067 | | | |
| 32 | | 44.522 | 45.084 | | | |
| 64 | | 45.044 | 45.087 | | | |
| 128 | | 45.176 | 45.088 | | | |
| 256 | | 45.209 | 45.088 | | | |
| 512 | | 45.217 | | | | |
| 1024 | | 45.220 | | | | |
| 2048 | | 45.220 | | | | |
| Converged | 44.527 | 45.220 | 45.088 | 45.132 | 45.088 | 45.102 |

**Table 3** Convergence of the tip displacement (mm) (C-F, Type C with $p$ = 5.0).

| Number of elements | DEB | DFS | DTS | MTS | PFTS-T | PFTS |
|---|---|---|---|---|---|---|
| 1 | 32.759 | 3.1784 | 28.417 | 34.133 | 37.095 | 37.232 |
| 2 | 35.286 | 10.107 | 35.144 | 36.567 | 37.095 | 37.232 |
| 4 | 35.917 | 22.214 | 36.717 | 37.165 | | |
| 8 | 36.075 | 31.709 | 37.021 | 37.309 | | |
| 16 | 36.115 | 35.503 | 37.075 | 37.342 | | |
| 32 | 36.125 | 36.597 | 37.089 | 37.350 | | |
| 64 | 36.127 | 36.881 | 37.093 | 37.352 | | |
| 128 | 36.128 | 36.953 | 37.094 | 37.352 | | |
| 256 | 36.128 | 36.971 | 37.094 | | | |
| 512 | | 36.976 | | | | |
| 1024 | | 36.977 | | | | |
| 2048 | | 36.977 | | | | |
| Converged | 36.128 | 36.977 | 37.094 | 37.352 | 37.095 | 37.232 |

**Table 4** Number of DoFs and relative computational time for convergence (C-F, Type B with $p$ = 5.0).

| | DEB | DFS | DTS | MTS | PFTS-T | PFTS |
|---|---|---|---|---|---|---|
| Number of DoFs | 3 | 6144 | 512 | 4 | 9 | 9 |
| $T_{\text{beam}}/T_{\text{DTS}}$ | 0.1043 | 82.826 | 1.0000 | 0.1087 | 0.1957 | 0.1957 |

**Table 5** Number of DoFs and relative computational time for convergence (C-F, Type C with $p$ = 5.0).

| | DEB | DFS | DTS | MTS | PFTS-T | PFTS |
|---|---|---|---|---|---|---|
| Number of DoFs | 384 | 6144 | 512 | 235 | 9 | 9 |
| $T_{\text{beam}}/T_{\text{DTS}}$ | 0.7304 | 82.826 | 1.0000 | 0.5478 | 0.1957 | 0.1957 |

The convergence test dictates the computational requirements: 128 elements for DEB and DTS, 1024 elements for DFS, 64 elements for MTS and a single element for both PFTS and PFTS-T. This discretization is implemented in the subsequent calculations. The displacement solutions for the three FG material models, under varying power-law index settings, are presented in **Table 6**. Especially, the solutions obtained from Q4 with a mesh of $m_x \times m_y = 401 \times 100$, where $m_x$ and $m_y$ denote the number of elements along $x$-axis and $y$-axis, respectively, are also presented in **Table 6** as the references, and the Average Relative Error (ARE) defined below is employed to assess the accuracy of the displacement solution:

$$\mathrm{ARE} = \frac{1}{n}\sum_{i=1}^{n}\left|\frac{d_{\mathrm{beam}} - d_{Q_4}}{d_{Q_4}}\right| \times 100\% \tag{133}$$

where $n$ represents the total number of settings for the material parameters ($n = 15$ in this study for different settings with three types of FG material model and five values of power-law index), $d_{Q_4}$ represents the displacement solutions obtained from Q4 and $d_{\mathrm{beam}}$ refers to the displacement solutions obtained from the beam element models. **Table 6** reveals a significant discrepancy between the displacement results of DEB and DFS and the reference solutions derived from Q4, attributable to an inadequate reflection of shear deformation. The computational accuracy of DTS and PFTS-T is essentially identical, given that PFTS-T employs traditional shear stiffness, thereby not satisfying the equilibrium relations. The solution accuracy of MTS and PFTS, which are both constructed based on the transverse shear stress derived from the differential equilibrium equation (Eq. (29)), is higher than that of DTS and PFTS-T, suggesting that adherence to the equilibrium condition enhances element accuracy. Although PFTS-T and PFTS are based on fundamentally similar formulas, the shear stiffness variation impacts their computational accuracy. Specifically, PFTS improves computational accuracy by introducing the modified shear stiffness to satisfy the equilibrium condition.

**Table 6** Comparison of the tip displacement solutions (mm) (C-F).

| Type | $p$ | DEB | DFS | DTS | MTS | PFTS-T | PFTS | Q4 |
|---|---|---|---|---|---|---|---|---|
| A | 0.0 | 13.158 | 13.569 | 13.563 | 13.567 | 13.564 | 13.564 | 13.569 |
| | 0.5 | 20.297 | 20.861 | 20.845 | 20.851 | 20.847 | 20.847 | 20.833 |
| | 1.0 | 26.398 | 27.092 | 27.081 | 27.091 | 27.084 | 27.086 | 27.073 |
| | 5.0 | 40.004 | 41.286 | 41.519 | 41.567 | 41.525 | 41.532 | 41.665 |
| | 10.0 | 43.919 | 45.508 | 45.791 | 45.827 | 45.798 | 45.809 | 45.960 |
| B | 0.0 | 13.158 | 13.569 | 13.563 | 13.567 | 13.564 | 13.564 | 13.569 |
| | 0.5 | 19.888 | 20.378 | 20.340 | 20.345 | 20.342 | 20.342 | 20.300 |
| | 1.0 | 25.528 | 26.072 | 26.012 | 26.019 | 26.015 | 26.016 | 25.960 |
| | 5.0 | 44.527 | 45.220 | 45.088 | 45.132 | 45.088 | 45.102 | 45.004 |
| | 10.0 | 49.891 | 50.631 | 50.466 | 50.565 | 50.471 | 50.495 | 50.431 |
| C | 0.0 | 26.230 | 26.774 | 26.705 | 26.759 | 26.708 | 26.738 | 26.714 |
| | 0.5 | 30.349 | 30.984 | 30.929 | 30.979 | 30.933 | 30.969 | 30.920 |
| | 1.0 | 32.630 | 33.324 | 33.292 | 33.368 | 33.297 | 33.363 | 33.324 |
| | 5.0 | 36.127 | 36.977 | 37.089 | 37.352 | 37.095 | 37.232 | 37.338 |
| | 10.0 | 36.405 | 37.300 | 37.476 | 37.811 | 37.482 | 37.631 | 37.753 |
| ARE(%) | | 2.53 | 0.42 | 0.21 | 0.16 | 0.20 | 0.17 | - |

An investigation on the distributions of axial normal stress and transverse shear stress is further conducted. For Type B and Type C material models with $p = 5.0$, the stress contours obtained from PFTS are plotted in **Fig. 7-Fig. 10**. In **Fig. 7-Fig. 10**, the stress distributions obtained from PFTS are compared with those obtained by DTS and Q4 at the three cross-sections of $x$ = 50mm, $x$ = 500mm and $x$ = 900mm.

**Fig. 7** and **Fig. 9** show that the axial normal stresses obtained from DTS, PFTS and Q4 are in good agreement. The slight difference in the results of DTS/PFTS and Q4 at the cross-section of $x$ = 900mm is mainly due to the differences in the loading modes at the free end: a concentrated load is applied in DTS/PFTS, while an equivalent distributed load is applied in Q4.

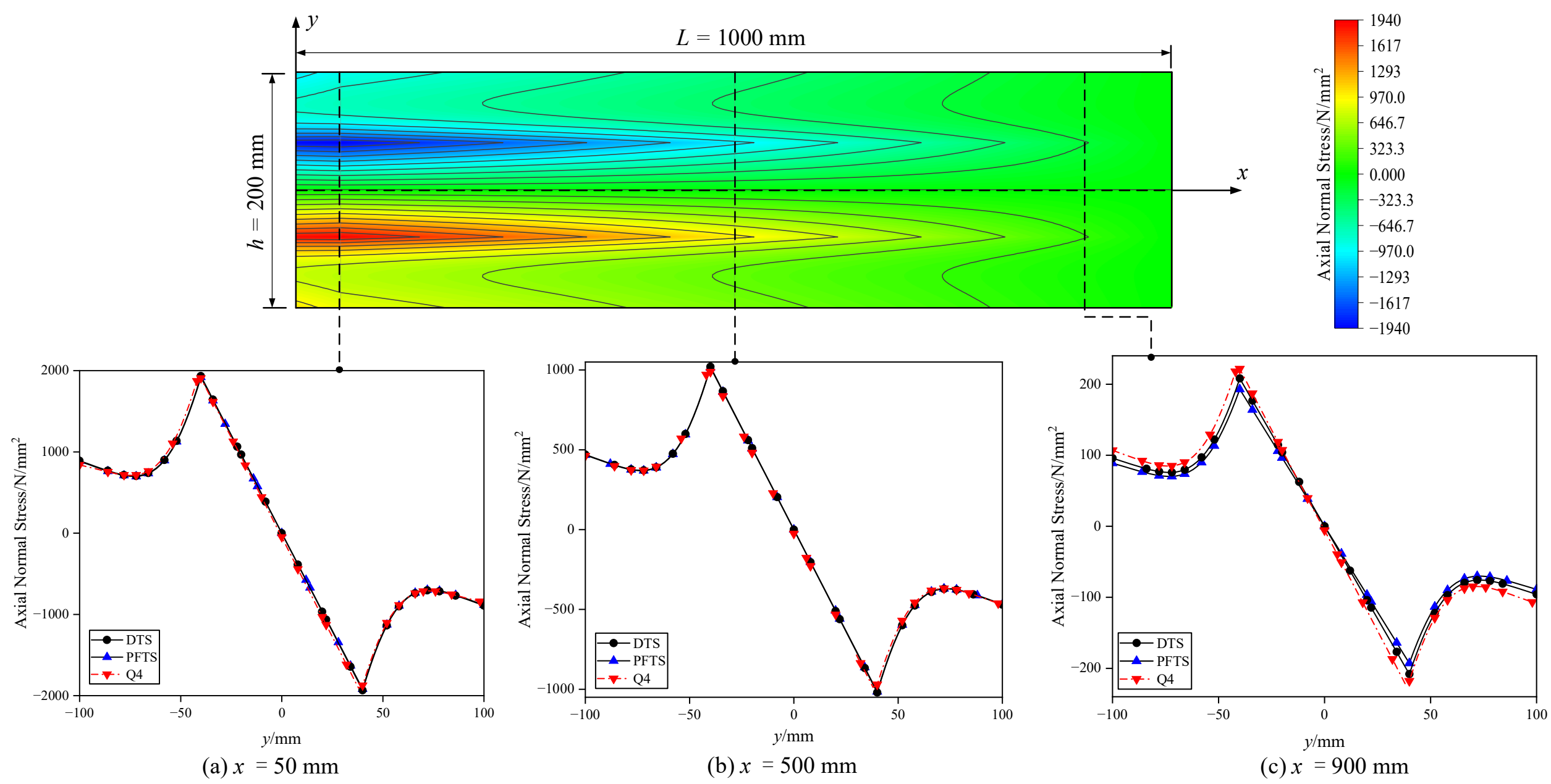

**Fig. 7**. Comparison of axial normal stress (C-F, Type B with $p$ = 5.0).

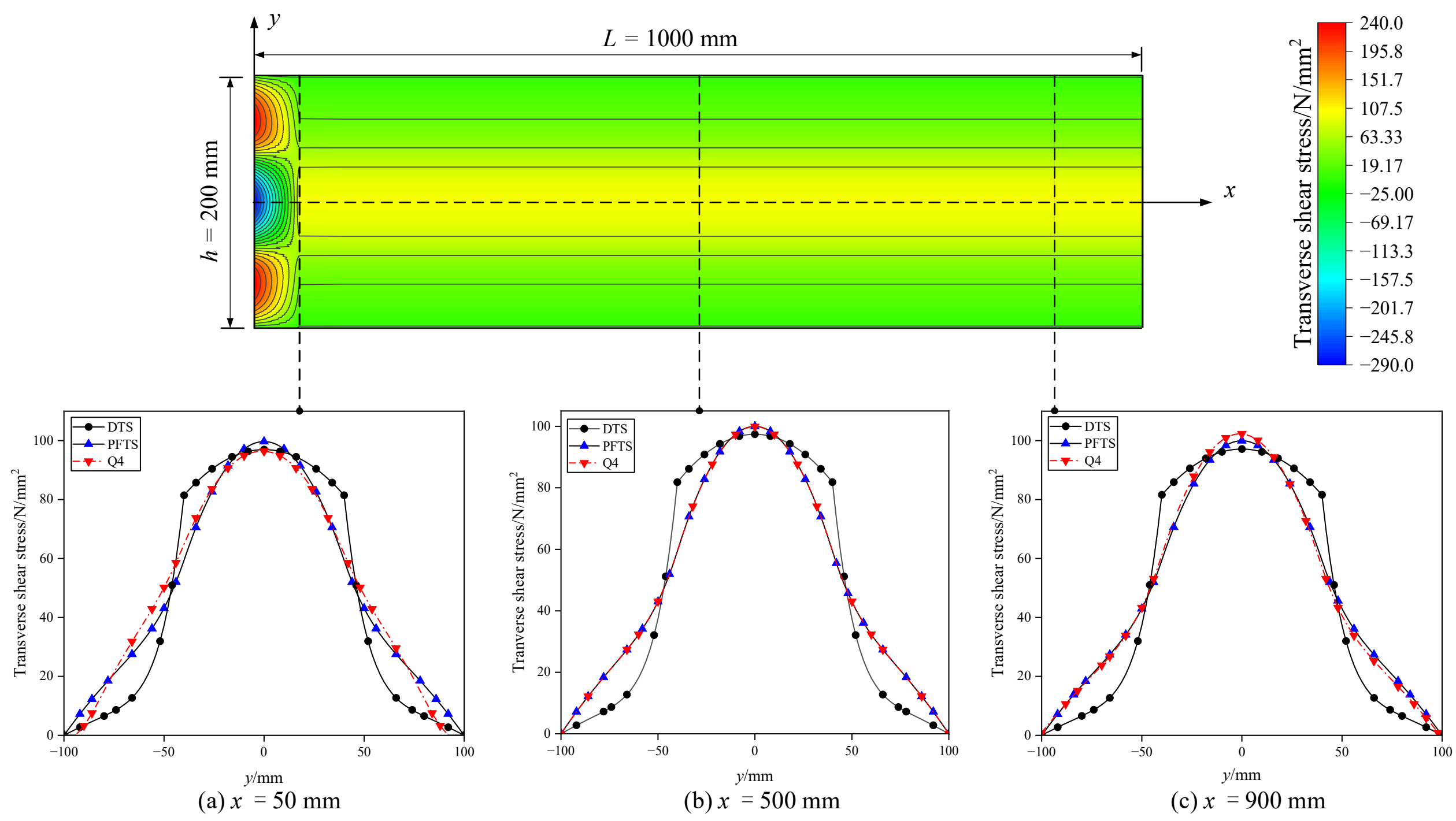

**Fig. 8**. Comparison of transverse shear stress (C-F, Type B with $p$ = 5.0).

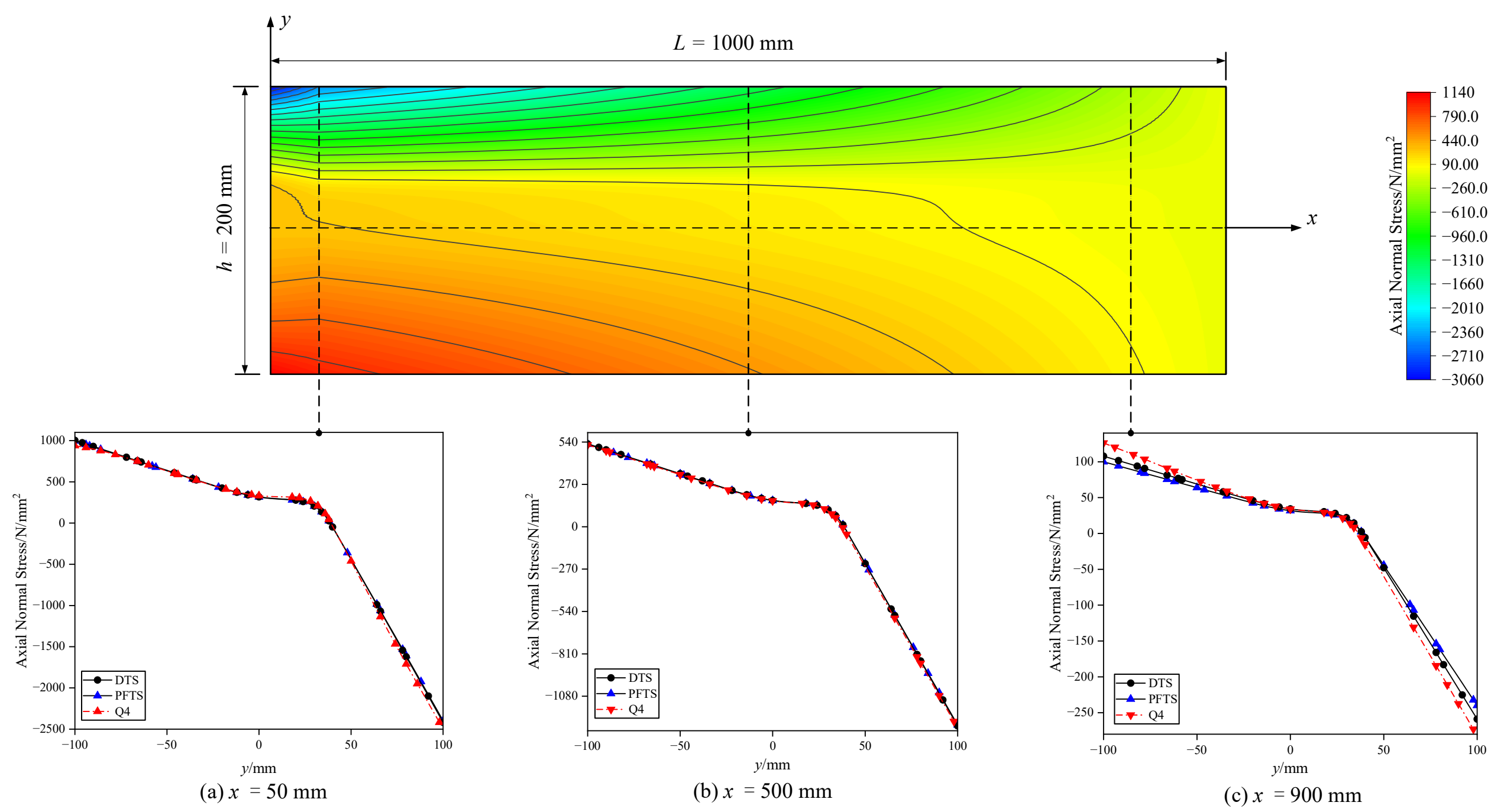

**Fig. 9**. Comparison of axial normal stress (C-F, Type C with $p$ = 5.0).

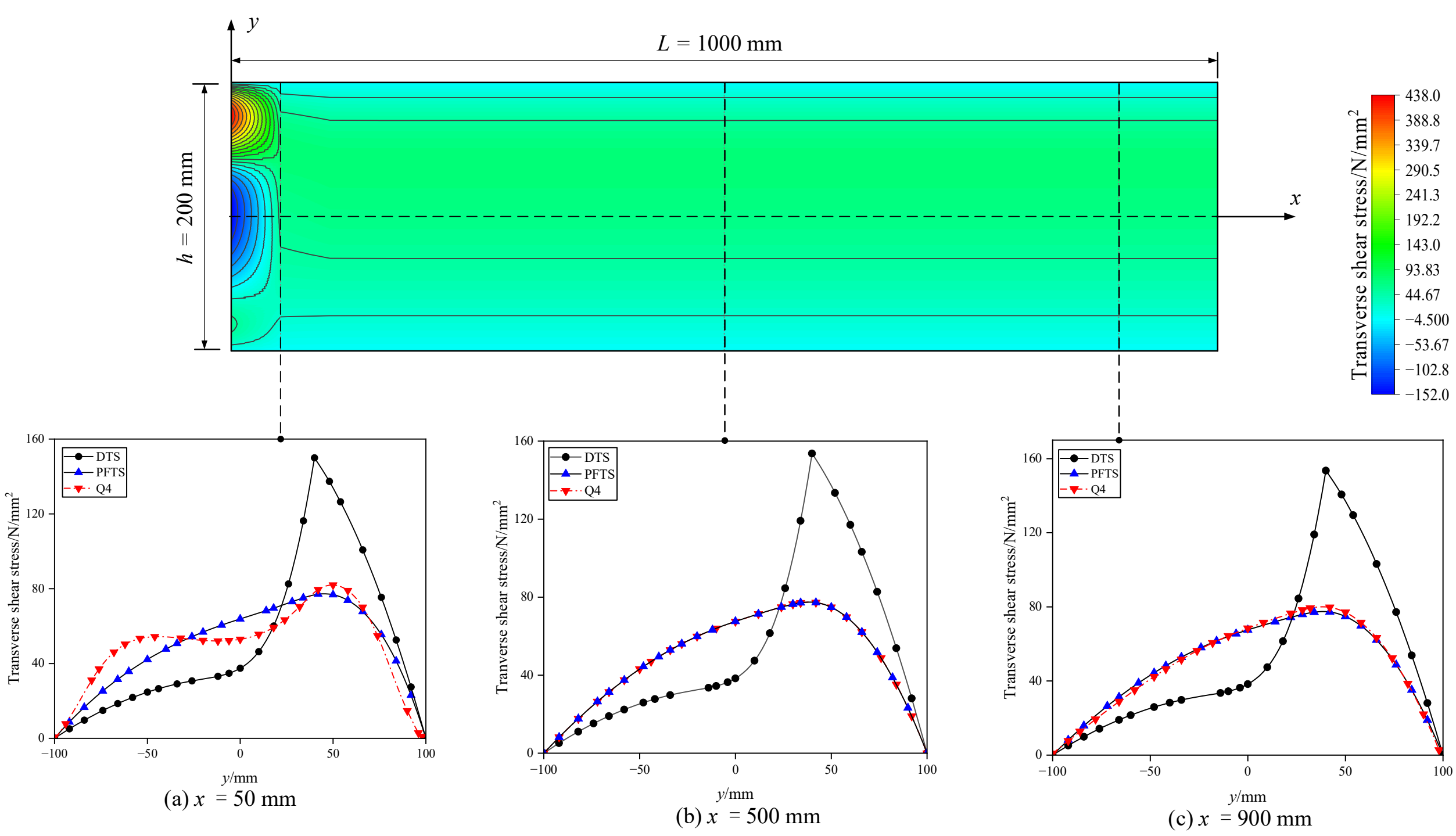

**Fig. 10**. Comparison of transverse shear stress (C-F, Type C with $p$ = 5.0).

**Fig. 8** and **Fig. 10** reveal that the shear stress distributions of DTS and PFTS differ significantly, which can be explained by their expressions and calculation modes for the transverse shear stress. In DTS, after acquiring the nodal displacements through finite element analysis, the generalized strain $\gamma_0(x)$ is obtained through displacement interpolation along with Eq. (8). The transverse shear strain is subsequently calculated using Eq. (4), and the corresponding transverse shear stress is determined through Eq. (13). In PFTS, both the nodal displacements of the structure and the internal force parameters for each element are determined simultaneously by solving the structural equation system. Subsequently, by using Eq. (85), $\boldsymbol{\tau}(x)=\{M_{w,x}(x) \quad M_{\theta,x}(x)\}^{\mathrm{T}}$ can be obtained from the internal force parameters in each element, which ultimately leads to the determination of the transverse shear stress, as described in Eq. (35).

As shown in **Fig. 8** and **Fig. 10**, the shear stress distributions of PFTS agree well with that of Q4 (except near the clamped end), indicating that the proposed element can capture the true shear stress distributions. Meanwhile, the distribution characteristics show that, even though the material properties vary non-smoothly along the beam's thickness, the distribution of transverse shear stress should be a smooth curve, which has been reflected in Eq. (35). Due to the difference in imposing constraints and stress field definitions between the beam element model (PFTS) and the plane 4-node element model (Q4), the shear stress distribution of PFTS still does not exactly match the Q4 results near the clamped end. However, the influence of this discrepancy on the outcomes is not substantial. For DTS, the shear stress distribution curve is derived from the geometric and constitutive relationships, as shown in Eqs. (8), (4) and (13). While the transverse shear strain, which is derived from the displacement function, manifests as a smooth curve, the resulting transverse shear stress exhibits a non-smooth distribution with abrupt changes at the interlayer junction, as indicated by **Fig. 10**. Therefore, the relative error of maximum transverse shear stress between DTS and Q4 is as high as 100%. This discrepancy signifies a substantial deviation from the true shear stress distribution and consequently impacts the precision of the displacement solutions.

For different settings of power-law index, the maximum values of transverse shear stress on the cross-section obtained by DTS and PFTS are listed in **Table 7**. Meanwhile, the results obtained from Q4 are used as the reference, and the Relative Error (RE) is presented to evaluate the accuracy of the beam elements. The data indicate that for different settings of power-law index, the transverse shear stresses obtained by PFTS are consistent with the references. For DTS, except in a few cases where acceptable accuracy can be obtained, the results in most cases have significant deviations, especially for the cases of Type C material with high power-law index. It is demonstrated that the modified HSBT established in this article can ensure the accuracy of transverse shear stress. Therefore, the performance of PFTS is excellent compared to the beam element models based on traditional higher-order shear deformation theory.

**Table 7** Comparison of the maximum transverse shear stress (N/mm$^2$) (C-F, $x$=500mm).

| Type | $p$ | DTS | | PFTS | | Q4 |
|---|---|---|---|---|---|---|
| | | Disp | RE(%) | Disp | RE(%) | |
| A | 0.0 | 75.025 | 0.12 | 75.000 | 0.09 | 74.935 |
| | 0.5 | 79.125 | 0.81 | 78.571 | 0.10 | 78.493 |
| | 1.0 | 82.483 | 3.25 | 79.937 | 0.06 | 79.889 |
| | 5.0 | 83.637 | 14.43 | 73.118 | 0.04 | 73.088 |
| | 10.0 | 66.515 | 6.08 | 70.854 | 0.05 | 70.820 |
| B | 0.0 | 75.025 | 0.12 | 75.000 | 0.09 | 74.935 |
| | 0.5 | 81.045 | 2.18 | 83.042 | 0.23 | 82.848 |
| | 1.0 | 85.237 | 3.55 | 88.534 | 0.18 | 88.376 |
| | 5.0 | 97.140 | 2.65 | 99.991 | 0.21 | 99.782 |
| | 10.0 | 100.97 | 0.31 | 100.88 | 0.23 | 100.65 |
| C | 0.0 | 87.238 | 2.2 | 89.371 | 0.06 | 89.317 |
| | 0.5 | 93.727 | 8.19 | 86.698 | 0.07 | 86.635 |
| | 1.0 | 104.89 | 24.31 | 84.444 | 0.08 | 84.380 |
| | 5.0 | 153.52 | 98.16 | 77.476 | 0.00 | 77.473 |
| | 10.0 | 170.99 | 125.32 | 75.921 | 0.04 | 75.887 |

4.3 FG beams under uniformly distributed load with both end supported

This section investigates a 2000mm-long beam subjected to a uniform load $q = 5000\,\mathrm{N/mm}$, as depicted in **Fig. 11**. Two support cases are examined: (1) Case A: simply-supported at both ends (S-S), and (2) Case B: clamped-clamped supported at both ends (C-C). Unlike the example in **Sec. 4.2**, the beam's shear force in this example varies along the beam axis due to the distributed load, thus examining the computational performance of the proposed beam model in the cases of non-uniform shear force distribution.

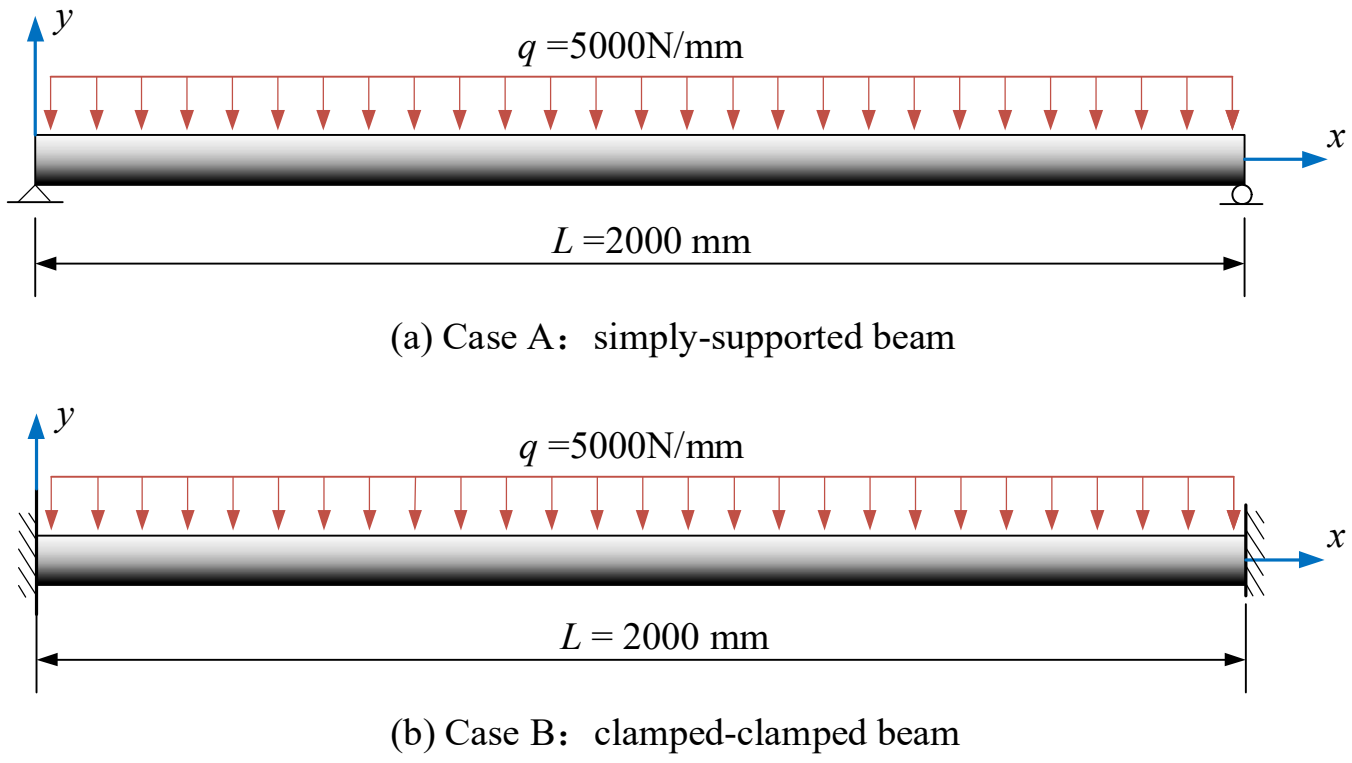

**Fig. 11**. Geometry of the both end supported beams.

Convergence analysis is performed on various beam element models using the FG beams with Type C material under the setting of $p$ = 5.0, and the results are presented in **Table 8** and **Table 9**. The convergence results indicate that for the beams with non-uniform shear force distribution along the beam axis, the beam element models (PFTS and PFTS-T) derived from the formulation presented in this paper require only a single element to achieve convergence, which can effectively circumvent the issue of discretization error. For the displacement-based beam elements (DEB, DFS and DTS) and the mixed beam element (MTS), due to the mismatch between the assumed polynomial forms and the actual displacement fields, it is necessary to refine the mesh to attain convergence. Furthermore, the solution efficiency of all beam elements is investigated. The number of DoFs as well as the relative computational time required for obtaining the converged solutions are listed in **Table 10** and **Table 11**. Similar to the example in **Sec. 4.2**, the computational efficiency of the element model proposed (PFTS and PFTS-T) has significant advantages compared to other beam element models, due to the requirement of only one element to achieve the converged solutions. Based on the convergence study, the number of elements for the subsequent studies can be determined as: 512 elements for DEB, DTS and MTS, 256 elements for DFS, and a single element for PFTS and PFTS-T.

**Table 8** Convergence of the mid-span displacement (mm) (S-S, Type C with $p$ = 5.0).

| Number of elements | DEB | DFS | DTS | MTS | PFTS-T | PFTS |
|---|---|---|---|---|---|---|
| 1 | - | - | - | - | 230.70 | 231.39 |
| 2 | 163.79 | 138.66 | 142.09 | 170.66 | 230.70 | 231.39 |
| 4 | 210.30 | 223.30 | 209.59 | 216.71 | | |
| 8 | 221.92 | 229.01 | 225.94 | 228.19 | | |
| 16 | 224.83 | 229.85 | 229.62 | 231.05 | | |
| 32 | 225.56 | 230.00 | 230.43 | 231.77 | | |
| 64 | 225.74 | 230.03 | 230.63 | 231.95 | | |
| 128 | 225.78 | 230.04 | 230.68 | 232.00 | | |
| 256 | 225.80 | 230.05 | 230.69 | 232.01 | | |
| 512 | 225.80 | 230.05 | 230.69 | 232.01 | | |
| Converged | 225.80 | 230.05 | 230.69 | 232.01 | 230.70 | 231.39 |

**Table 9** Convergence of the mid-span displacement (mm) (C-C, Type C with $p$ = 5.0).

| Number of elements | DEB | DFS | DTS | MTS | PFTS-T | PFTS |
|---|---|---|---|---|---|---|
| 1 | - | - | - | - | 49.929 | 50.606 |
| 2 | 28.313 | 18.689 | 142.09 | 34.882 | 49.929 | 50.606 |
| 4 | 40.948 | 45.488 | 40.237 | 47.356 | | |
| 8 | 44.107 | 49.076 | 48.085 | 50.317 | | |
| 16 | 44.897 | 49.384 | 49.549 | 50.998 | | |
| 32 | 45.094 | 49.405 | 49.815 | 51.145 | | |
| 64 | 45.143 | 49.407 | 49.894 | 51.178 | | |
| 128 | 45.156 | 49.406 | 49.920 | 51.185 | | |
| 256 | 45.159 | 49.406 | 49.927 | 51.187 | | |
| 512 | 45.160 | | 49.929 | 51.188 | | |
| 1024 | 45.160 | | 49.929 | 51.188 | | |
| Converged | 45.160 | 49.406 | 49.929 | 51.188 | 49.929 | 50.606 |

**Table 10** Number of DoFs and relative computational time for convergence (S-S, Type C with $p$ = 5.0).

| | DEB | DFS | DTS | MTS | PFTS-T | PFTS |
|---|---|---|---|---|---|---|
| Number of DoFs | 768 | 1536 | 1025 | 1025 | 9 | 9 |
| $T_{\text{beam}}/T_{\text{DTS}}$ | 0.7447 | 2.0662 | 1.0000 | 1.1891 | 0.1064 | 0.1064 |

**Table 11** Number of DoFs and relative computational time for convergence (C-C, Type C with $p$ = 5.0).

| | DEB | DFS | DTS | MTS | PFTS-T | PFTS |
|---|---|---|---|---|---|---|
| Number of DoFs | 1533 | 765 | 2044 | 2044 | 9 | 9 |
| $T_{\text{beam}}/T_{\text{DTS}}$ | 0.7000 | 0.2856 | 1.0000 | 1.0976 | 0.0360 | 0.0360 |

Considering two boundary conditions (S-S and C-C), the vertical displacements at the mid-span obtained by different beam elements for various material models are listed in **Table 12** and **Table 13**. Meanwhile, the solutions obtained from Q4 with a mesh of $m_x \times m_y = 401\times100$ are presented as the references. The findings from **Table 12** and **Table 13** reveal that PFTS demonstrates superior precision in comparison to DEB, DFS, and DTS. This outcome aligns with the observations from the numerical example of the cantilever, and further underscores that adherence to the equilibrium condition can markedly enhance the accuracy of the element, particularly in scenarios where the shear force varies along the beam axis. As shown in **Table 12** and **Table 13**, the solution accuracy of DEB is very low because it does not consider shear deformation. For cases of Type B material model, the solution accuracy of shear beam elements is similar, while for cases of Type C material, PFTS exhibits significantly higher computational accuracy. This study validates the reliability of the proposed theory and beam element in solving the displacement for beams with non-uniform shear force distribution.

**Fig. 12** and **Fig. 13** present the contours of transverse shear stress obtained from PFTS for S-S beam with Type B and Type C material models under $p$ = 5.0, respectively. In **Fig. 12** and **Fig. 13**, the stress distributions obtained from PFTS are compared with those obtained by DTS and Q4 at the five cross-sections of $x$ = 50mm, $x$ = 500mm, $x$ = 1000mm, $x$ = 1500mm and $x$ = 1800mm. Due to the variation of shear force along the beam axis, the shear stress exhibits variations in magnitude across different cross-sections, despite the resemblance in distribution shapes. It is noteworthy that, for the beams under a uniform load, both the shear force and shear stress are null at the middle cross-section ($x$ = 1000mm), while the maximum shear force occurs at the two supports, corroborating the results illustrated in the figures. The methods for working out transverse shear stress in DTS and PFTS, as well as the formulas used, have been explained in **Sec. 4.2** and will not be repeated here. As discerned from the outcomes presented in **Fig. 12** and **Fig. 13**, the stress distributions procured by PFTS align with those of Q4. While the magnitude of transverse shear stress varies at different cross-sections, their shapes are similar and smoothly changes along the thickness of the cross-section, which is consistent with the transverse shear stress expression provided in Eq. (35). This observation underscores that the element model based on the stress equilibrium condition delineated in this study is equally applicable to scenarios where the shear undergoes variations along the beam axis. Conversely, DTS exhibits smooth variations along the cross-section in the transverse shear strain, while the transverse shear stress cannot be kept smooth. Therefore, DTS falls short in achieving a reasonable shear stress distribution, thereby compromising its computational precision.

**Table 12** Comparison of the mid-span displacement solutions (mm) (S-S).

| Type | $p$ | DEB | DFS | DTS | MTS | PFTS-T | PFTS | Q4 |
|---|---|---|---|---|---|---|---|---|
| A | 0.0 | 82.237 | 84.290 | 84.289 | 84.289 | 84.289 | 84.289 | 84.323 |
| | 0.5 | 126.86 | 129.68 | 129.63 | 129.63 | 129.63 | 129.64 | 129.51 |
| | 1.0 | 164.99 | 168.46 | 168.45 | 168.47 | 168.45 | 168.47 | 168.31 |
| | 5.0 | 250.02 | 256.43 | 257.73 | 257.90 | 257.73 | 257.77 | 258.67 |
| | 10.0 | 274.49 | 282.44 | 284.01 | 284.09 | 284.01 | 284.07 | 285.18 |
| B | 0.0 | 82.237 | 84.290 | 84.289 | 84.289 | 84.289 | 84.289 | 84.323 |
| | 0.5 | 124.30 | 126.75 | 126.59 | 126.59 | 126.59 | 126.59 | 126.89 |
| | 1.0 | 159.55 | 162.27 | 162.00 | 162.01 | 162.00 | 162.01 | 162.32 |
| | 5.0 | 278.29 | 281.76 | 281.12 | 281.32 | 281.12 | 281.19 | 281.46 |
| | 10.0 | 311.82 | 315.52 | 314.74 | 315.19 | 314.74 | 314.86 | 315.36 |
| C | 0.0 | 163.94 | 166.66 | 166.35 | 166.58 | 166.35 | 166.50 | 166.92 |
| | 0.5 | 189.68 | 192.85 | 192.63 | 192.84 | 192.63 | 192.81 | 193.24 |
| | 1.0 | 203.94 | 207.41 | 207.31 | 207.64 | 207.31 | 207.64 | 208.25 |
| | 5.0 | 225.80 | 230.05 | 230.69 | 232.01 | 230.70 | 231.39 | 232.59 |
| | 10.0 | 227.53 | 232.01 | 232.98 | 234.68 | 232.99 | 233.75 | 235.57 |
| RE(%) | | 2.27 | 0.39 | 0.32 | 0.19 | 0.32 | 0.25 | - |

**Table 13** Comparison of the mid-span displacement solutions (mm) (C-C).

| Type | $p$ | DEB | DFS | DTS | MTS | PFTS-T | PFTS | Q4 |
|---|---|---|---|---|---|---|---|---|
| A | 0.0 | 16.447 | 18.500 | 18.454 | 18.485 | 18.454 | 18.454 | 18.389 |
| | 0.5 | 25.372 | 28.191 | 28.089 | 28.130 | 28.089 | 28.090 | 27.970 |
| | 1.0 | 32.998 | 36.464 | 36.387 | 36.450 | 36.387 | 36.399 | 36.247 |
| | 5.0 | 50.005 | 56.416 | 57.504 | 57.751 | 57.505 | 57.542 | 57.461 |
| | 10.0 | 54.899 | 62.844 | 64.162 | 64.376 | 64.163 | 64.218 | 64.050 |
| B | 0.0 | 16.447 | 18.500 | 18.454 | 18.485 | 18.454 | 18.454 | 18.389 |
| | 0.5 | 24.86 | 27.312 | 27.109 | 27.140 | 27.109 | 27.110 | 26.968 |
| | 1.0 | 31.911 | 34.628 | 34.321 | 34.357 | 34.321 | 34.327 | 34.153 |
| | 5.0 | 55.659 | 59.125 | 58.444 | 58.683 | 58.444 | 58.513 | 58.344 |
| | 10.0 | 62.364 | 66.063 | 65.241 | 65.729 | 65.241 | 65.359 | 65.373 |
| C | 0.0 | 32.788 | 35.506 | 35.150 | 35.430 | 35.151 | 35.301 | 35.229 |
| | 0.5 | 37.936 | 41.111 | 40.825 | 41.081 | 40.825 | 41.002 | 40.831 |
| | 1.0 | 40.788 | 44.255 | 44.081 | 44.461 | 44.081 | 44.407 | 44.198 |
| | 5.0 | 45.160 | 49.407 | 49.929 | 51.188 | 49.929 | 50.606 | 50.903 |
| | 10.0 | 45.506 | 49.982 | 50.816 | 52.406 | 50.816 | 51.550 | 51.960 |
| RE(%) | | 9.04 | 1.31 | 0.52 | 0.58 | 0.52 | 0.38 | - |

Similar to **Fig. 12** and **Fig. 13**, **Fig. 14** illustrates the contour of transverse shear stress obtained from PFTS for the C-C beam with Type C material model under $p$ = 5.0. In **Fig. 14**, the shear stress results obtained by DTS, PFTS and Q4 at the five cross-sections of $x$ = 50mm, $x$ = 500mm, $x$ = 1000mm, $x$ = 1500mm and $x$ = 1950mm are compared. As shown in **Fig. 14**, except for a slight mismatch near the clamped end, the transverse shear stress obtained by PFTS is almost identical to  that of Q4. Note that the transverse shear stresses obtained from PFTS and Q4 vary smoothly along the thickness of the cross-section, which is consistent with the transverse shear stress expression (Eq. (35)). The mismatch in shear stress distribution between PFTS and Q4 is due to the difference in imposing constraints and stress field definitions between the beam element model (PFTS) and the plane 4-node element model (Q4).  **Fig. 14** illustrates that the transverse shear stress obtained by DTS differs significantly from that of Q4. Being similar to the first example, the relative error of maximum transverse shear stress between DTS and Q4 is as high as approximately 100%. The maximum values of transverse shear stress of the cross-section obtained by DTS and PFTS under different settings of power-law index are examined, as shown in **Table 14**, where the results obtained from Q4 are used as the reference. The data indicate that DTS produces significant relative errors in most cases, while PFTS achieves satisfactory solution accuracy. These results indicte that the modified HSBT has greatly improved the accuracy of stress solution compared to the traditional HSBT.

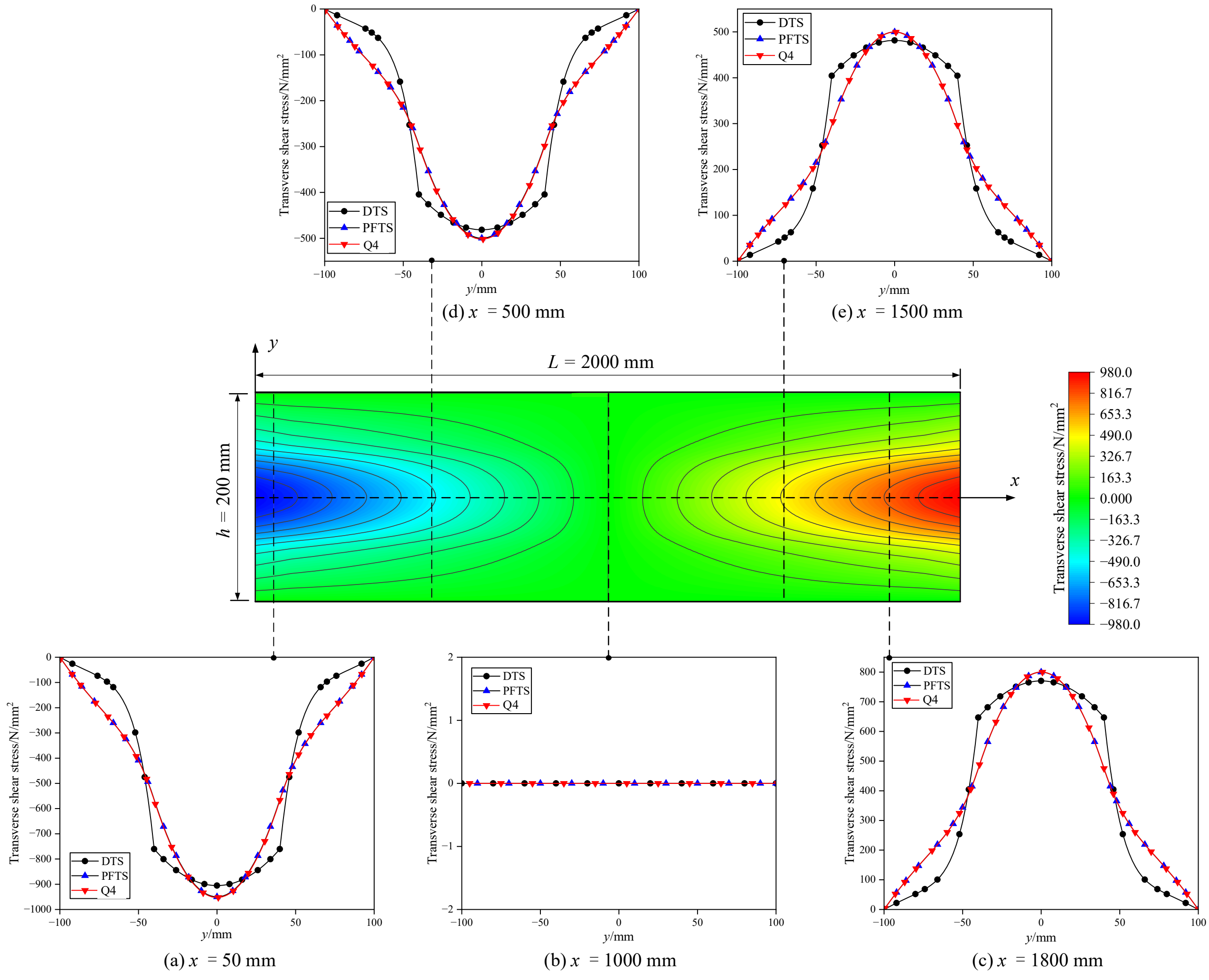

**Fig. 12**. Comparison of transverse shear stress (S-S, Type B with $p$ = 5.0).

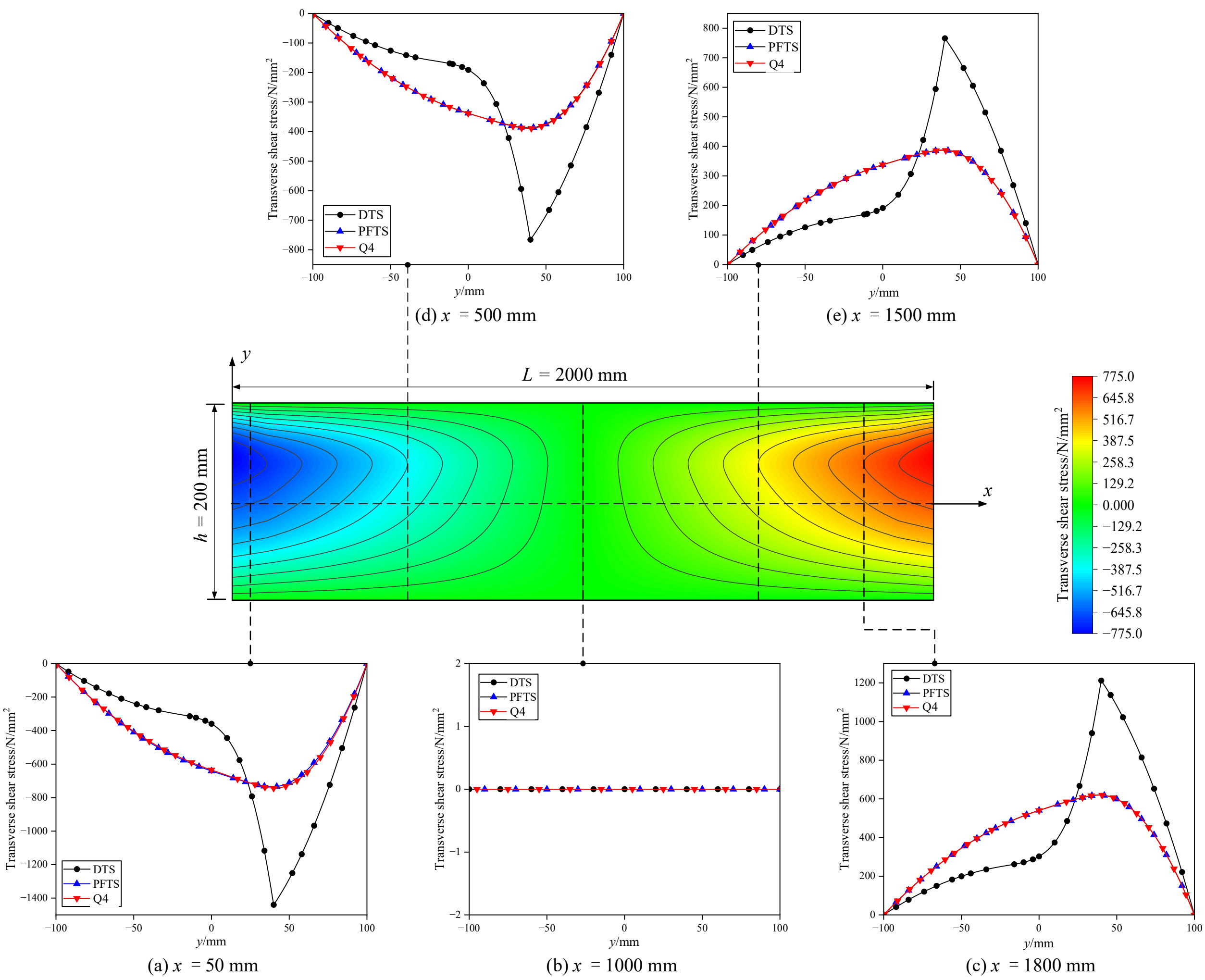

**Fig. 13**. Comparison of transverse shear stress (S-S, Type C with $p$ = 5.0).

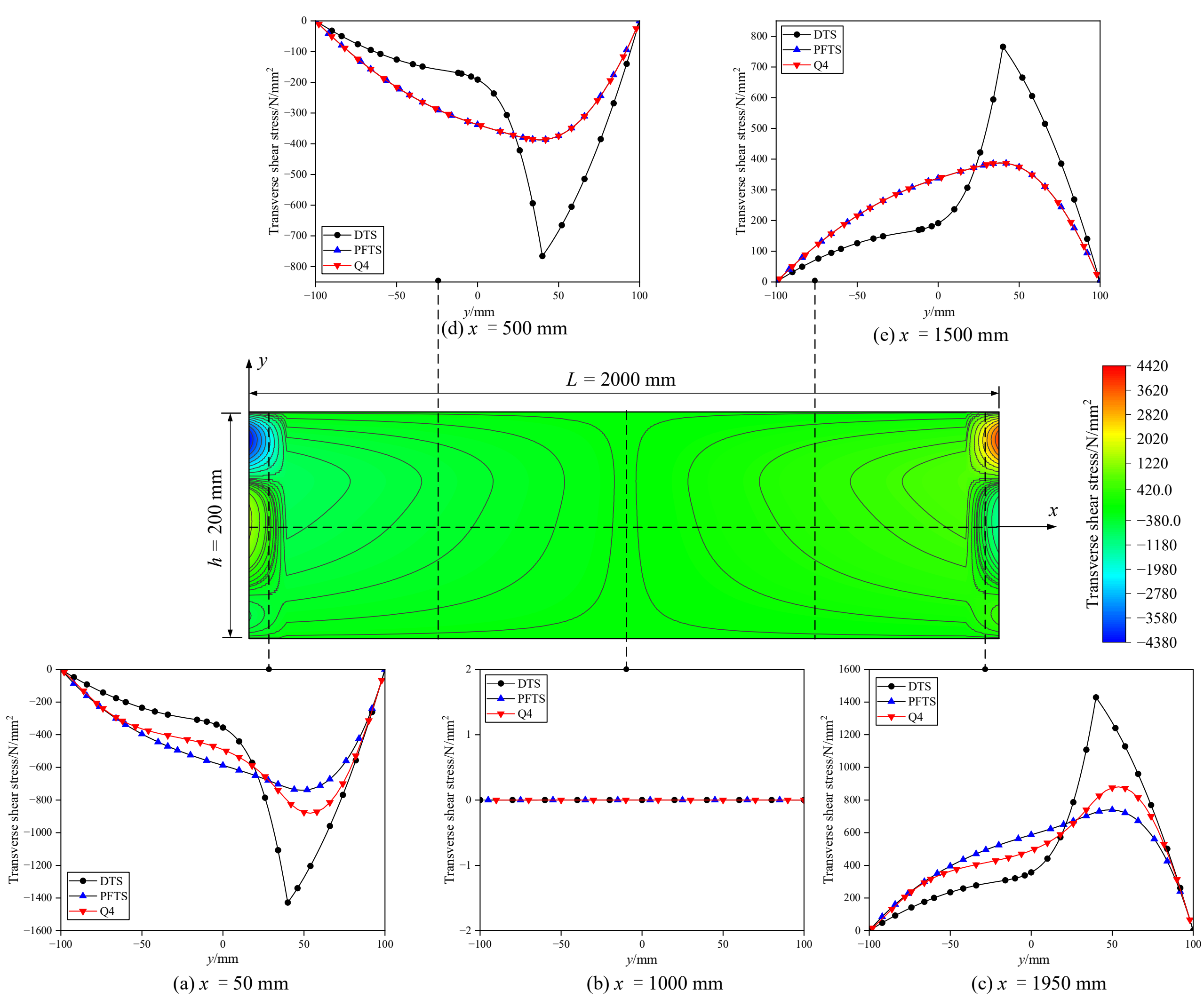

**Fig. 14**. Comparison of transverse shear stress (C-C, Type C with $p$ = 5.0).

**Table 14** Comparison of the maximum transverse shear stress (N/mm$^2$) (C-C, $x$=1500mm).

| Type | $p$ | DTS | | PFTS | | Q4 |
|---|---|---|---|---|---|---|
| | | Disp | RE(%) | Disp | RE(%) | |
| A | 0.0 | 373.99 | 0.18 | 375.00 | 0.09 | 74.935 |
| | 0.5 | 394.34 | 0.48 | 392.85 | 0.10 | 78.493 |
| | 1.0 | 411.17 | 2.93 | 399.68 | 0.06 | 79.889 |
| | 5.0 | 417.44 | 13.2 | 365.59 | 0.04 | 73.088 |
| | 10.0 | 332.02 | 6.24 | 354.27 | 0.05 | 70.820 |
| B | 0.0 | 373.99 | 0.18 | 375.00 | 0.09 | 74.935 |
| | 0.5 | 403.34 | 2.63 | 415.20 | 0.23 | 82.848 |
| | 1.0 | 423.65 | 4.13 | 442.66 | 0.18 | 88.376 |
| | 5.0 | 481.46 | 3.50 | 499.94 | 0.21 | 99.782 |
| | 10.0 | 500.25 | 0.60 | 504.41 | 0.23 | 100.65 |
| C | 0.0 | 434.80 | 2.64 | 446.85 | 0.06 | 89.317 |
| | 0.5 | 466.94 | 7.80 | 433.49 | 0.07 | 86.635 |
| | 1.0 | 522.75 | 23.9 | 422.22 | 0.08 | 84.380 |
| | 5.0 | 765.87 | 97.7 | 387.38 | 0.00 | 77.473 |
| | 10.0 | 853.19 | 124.9 | 379.60 | 0.04 | 75.887 |

An extra investigation is conducted for the influence of the beam's slenderness ratio on the solution accuracy of the beam element models. As shown in **Fig. 15**, the C-C beam under a uniform load is considered. The beam's cross-section is $b \times h = 200\text{mm} \times 500\text{mm}$, and the beam's length is set to $L = 1000\text{mm}/2000\text{mm}/3000\text{mm}/4000\text{mm}/5000\text{mm}$, respectively. The beam element models selected in this analysis include: DFS, DTS, PFTS and Quasi-3D [70], where Quasi-3D represents the beam element based on quasi-3D theory [70, 71]. The displacement solutions obtained from Q4 are employed as the references. For the beams with Type C material model under $p$ = 5.0, the vertical displacements at the mid-span obtained by different beam element models as well as their relative errors are listed in **Table 15**. It can be seen from **Table 15** that, as the slenderness ratio increases, the displacement solutions of DFS, DTS and PFTS gradually tend towards that of Q4, indicating that the effect of shear deformation gradually decreases with the increase of slenderness ratio. For Quasi-3D , which includes both transverse normal strain and transverse normal stress, can achieve relatively ideal accuracy for the cases of low slenderness ratio. However, due to its inability to accurately describe the distribution of transverse normal stress, there is a significant difference between the results obtained by Quasi-3D and Q4 as the slenderness ratio increases. This investigation indicates that the beam element model proposed in this paper is applicable for FG beams under different slenderness ratios and achieves ideal results.

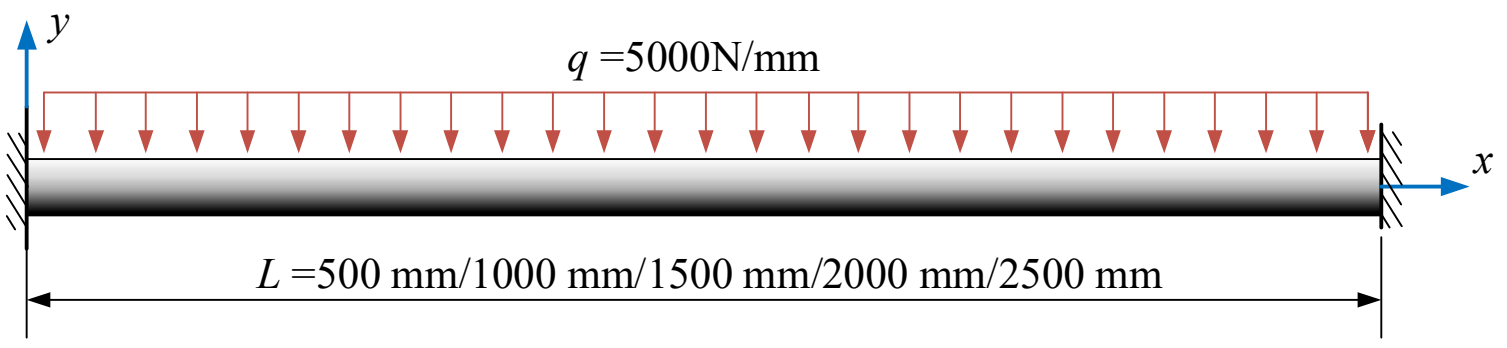

**Fig. 15**. The C-C beam under different settings of length.

**Table 15** Comparison of the mid-span displacements under different slenderness (C-C, Type C, $p$ = 5.0).

| $L$/mm | $L/h$ | DFS | | DTS | | Quasi-3D | | PFTS | | Q4 |
|---|---|---|---|---|---|---|---|---|---|---|
| | | Disp/mm | RE/% | Disp/mm | RE/% | Disp/mm | RE/% | Disp/mm | RE/% | |
| 1000 | 5 | 3.8846 | 8.76 | 3.9844 | 6.41 | 4.1557 | 2.39 | 4.1464 | 2.61 | 4.2574 |
| 2000 | 10 | 49.407 | 2.94 | 49.929 | 1.91 | 49.503 | 2.75 | 50.606 | 0.58 | 50.903 |
| 3000 | 15 | 238.18 | 1.09 | 239.44 | 0.57 | 234.16 | 2.76 | 240.99 | 0.07 | 240.81 |
| 4000 | 20 | 739.54 | 0.62 | 741.87 | 0.31 | 721.73 | 3.01 | 744.65 | 0.07 | 744.14 |
| 5000 | 25 | 1790.6 | 0.39 | 1794.3 | 0.19 | 1741.3 | 3.14 | 1798.7 | 0.06 | 1797.7 |

An additional investigation is conducted on the solution accuracy of PFTS in the beam's coupling behaviour of axial deformation and bending deformation. As shown in **Fig. 16**, the S-S beam under an axial force is considered, with a cross-section of $b \times h = 200\text{mm} \times 500\text{mm}$ and the length of 2000mm. Two asymmetric material models including Type A and Type B models are considered in this analysis, and the axial force is set to $F = 5 \times 10^7\,\text{N}$. For the beams with asymmetric material distribution, there is coupling between axial deformation and bending deformation, so the beam will undergo bending deformation under axial force. Therefore, this study examines the solution accuracy of beam elements based on the vertical displacement at the beam's mid-span. The displacements obtained by DFS, DTS, PFTS and Q4 as well as the average relative errors (ARE, see Eq. (133) with $n = 10$) of the beam elements are presented in **Table 16**. The results indicate that DFS, DTS and PFTS can take into account the coupling effect of axial deformation and bending deformation, and obtain relatively accurate displacement solutions. It is worth mentioning that the present numerical example does not involve shear deformation, so the results obtained by DTS and PFTS are basically consistent.

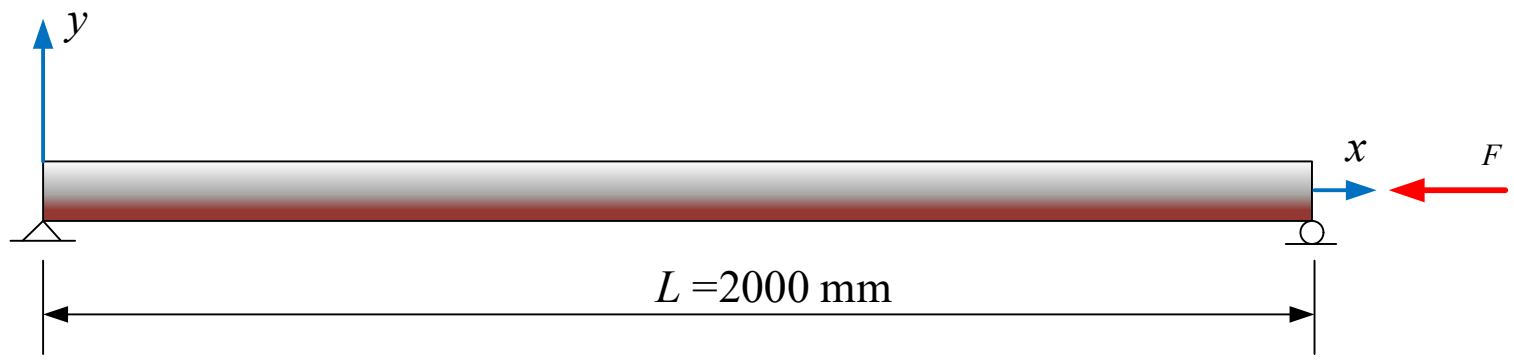

**Fig. 16**. The S-S beam under axial force.

**Table 16** Comparison of the mid-span displacements under axial force.

| Type | *p* | DFS | DTS | PFTS | Q4 |
|---|---|---|---|---|---|
| A | 0.0 | 0.0000 | 0.0000 | 0.0000 | 0.0000 |
| | 0.5 | 45.483 | 45.495 | 45.495 | 45.486 |
| | 1.0 | 90.927 | 90.927 | 90.927 | 91.107 |
| | 5.0 | 182.01 | 182.02 | 182.02 | 182.95 |
| | 10.0 | 157.58 | 157.74 | 157.74 | 156.84 |
| C | 0.0 | 89.247 | 89.248 | 89.248 | 88.989 |
| | 0.5 | 132.88 | 132.83 | 132.82 | 133.19 |
| | 1.0 | 159.60 | 159.50 | 159.49 | 160.75 |
| | 5.0 | 209.50 | 209.32 | 209.30 | 212.39 |
| | 10.0 | 215.76 | 215.59 | 215.56 | 217.24 |
| ARE(%) | | 0.49 | 0.54 | 0.54 | - |

## 5 Conclusions

This paper presents a modified higher-order shear deformation beam theory and a corresponding beam finite element model aimed at achieving precise analysis of functionally graded beams. In the modified theory, the distribution of transverse shear stress across the thickness of the beam is derived from the differential equilibrium equation, leading to the new formulation of the shear stiffness that accounts for the rational distribution of shear stress. Building upon the modified theory, expressions of internal forces are obtained from the closed-form solutions of the differential equilibrium equations associated with higher-order shear beams. A novel force-based higher-order beam element model is proposed, wherein the equation system for the beam element is established based on the equilibrium conditions at the element boundaries and the compatibility condition within the element. Numerical examples demonstrate that the proposed beam element offers significant advantages in both solution accuracy and computational efficiency for the static analysis of functionally graded beams. Additionally, several conclusions can be drawn from this study:

(1) The modified higher-order shear deformation theory, developed through the introduction of a modified shear stiffness, ensures the accuracy of the transverse shear stress distribution. Consequently, the beam element model based on this modified theory is capable of producing precise transverse shear stress, thereby enhancing solution accuracy.

(2) The force-based beam element model proposed in this study, with its equation system constructed upon the equilibrium relationships at the nodes and the deformation compatibility within the element, is demonstrated to be both feasible and reliable.

(3) The closed-form solutions for internal forces, derived from the differential equilibrium equations of the beam, enable the proposed higher-order shear beam element to reduce the effects of discretization errors, thereby ensuring the precision and stability of the solutions.

Although the equation system for the proposed beam element is formulated differently from that of conventional beam elements, it is still possible to derive the stiffness equation by condensing the degrees of freedom within the element. Under the premise of geometric linearity, the relationship between the displacement field and the internal force field of the element is relatively stable. Consequently, referring to the methods of exact finite element [60, 62], the proposed force-based beam model has the potential to be extended to the buckling and free vibration analyses of functionally graded beams, with the primary challenge being the derivation of internal force expressions for the corresponding nonlinear eigenvalue problem. In the case of geometrically nonlinear analysis, the expressions for internal forces become complex due to the necessity of accounting for additional nonlinear factors, which poses a significant challenge. The development of the force-based higher-order beam element to buckling analysis, free vibration analysis and geometrically nonlinear analysis represents a subject that warrants further investigation.

## Acknowledgments

The project is funded by the National Natural Science Foundation of China (Grant No. 52178209, Grant No. 51878299) and Guangdong Basic and Applied Basic Research Foundation, China (Grant No. 2021A1515012280, Grant No. 2020A1515010611).